\documentclass[journal,draftcls,onecolumn,12pt,twoside]{IEEEtran}
\usepackage{graphicx,multirow}
\usepackage{epsfig}
\usepackage{algorithm}
\usepackage{url}
\usepackage{algorithmic}
\usepackage{cite}
\usepackage{amssymb}
\usepackage{amsthm}
\usepackage{amsmath,subfigure}
\usepackage{color}
\usepackage{bbm}
\usepackage{cite}
\usepackage{epstopdf}
\usepackage[table,xcdraw]{xcolor}
\newtheorem{theorem}{\textbf{\text{Theorem}}}
\newtheorem{lemma}{\textbf{\text{Lemma}}}

\newtheorem{corollary}{Corollary}

\newtheorem{approximation}{Approximation}
\newtheorem{remark}{Remark}
\usepackage{collcell, datatool}
\usepackage{pifont}

\definecolor{lightgray}{gray}{0.9}
\usepackage{algorithm} 
\usepackage{algorithmic}  
\usepackage[algo2e]{algorithm2e} 
\renewcommand{\algorithmiccomment}[1]{}
\newcommand{\RN}[1]{%
  \textup{\uppercase\expandafter{\romannumeral#1}}%
}

\newcommand*{\mathcolor}{}
\def\mathcolor#1#{\mathcoloraux{#1}}
\newcommand*{\mathcoloraux}[3]{%
	\protect\leavevmode
	\begingroup
	\color#1{#2}#3%
	\endgroup
}
\begin{document}

\title{Spatiotemporal Stochastic Modeling of IoT Enabled Cellular Networks: Scalability and Stability Analysis}
\author{Mohammad Gharbieh, Hesham ElSawy, Ahmed Bader,\\ and Mohamed-Slim Alouini


\thanks{ The authors are with the Electrical Engineering Program, Computer, Electrical, and Mathematical Sciences and Engineering (CEMSE) Division, King Abdullah University of Science and Technology (KAUST), Thuwal, Saudi Arabia; e-mails: \{mohammad.gharbieh, hesham.elsawy, ahmed.bader, slim.alouini\}@kaust.edu.sa.}
}

\maketitle
\thispagestyle{plain}
\pagestyle{plain}

\vspace{-15mm}
\begin{abstract}

The Internet of Things (IoT) is large-scale by nature, which is manifested by the massive number of connected devices as well as their vast spatial existence. Cellular networks, which provide ubiquitous, reliable, and efficient wireless access, will play fundamental rule in delivering the first-mile access for the data tsunami to be generated by the IoT. However, cellular networks may have scalability problems to provide uplink connectivity to massive numbers of connected things. To characterize the scalability of cellular uplink in the context of IoT networks, this paper develops a traffic-aware spatiotemporal mathematical model for IoT devices supported by  cellular uplink connectivity. The developed model is based on stochastic geometry and queueing theory to account for the traffic requirement per IoT device, the different transmission strategies, and the mutual interference between the IoT devices. To this end, the developed model is utilized to characterize the extent to which cellular networks can accommodate IoT traffic as well as to assess and compare three different transmission strategies that incorporate a combination of transmission persistency, backoff, and power-ramping. The analysis and the results clearly illustrate the scalability problem imposed by IoT on cellular network and offer insights into effective scenarios for each transmission strategy.

\begin{IEEEkeywords} 
\vspace{-2mm}
IoT, LTE cellular networks, random access, stability, stochastic geometry, queueing theory, interacting queues.
\end{IEEEkeywords}

\end{abstract}

\vspace{-10mm}
\section{Introduction}

The Internet of Things (IoT) is expected to involve massive numbers of sensors, smart physical objects, machines, vehicles, and devices that require occasional data exchange and wireless Internet access~\cite{IOT1, NB-IOT}. Based on the IoT concept, a plethora of applications are emerging in possibly every industrial and vertical market, including vehicular communication, proximity services, wearable devices, autonomous driving, public safety, massive sensors support, and smart cities applications~\cite{IOT1, NB-IOT}.  Cellular networks are expected to play a fundamental role to provide first mile connectivity for a big sector of the IoT~\cite{IOT2,IoT22}. However, cellular networks are mainly designed to handle massive downlink traffic demands and the IoT is pushing the traffic to the uplink direction. Several studies report scalability issues in cellular networks for supporting massive uplink devices due to the random access based uplink scheduling~\cite{IOT2, RACH1, RACH2}.   
Consequently, the next evolution of cellular networks is not only envisioned to offer tangible performance improvement in terms of data rate, network capacity, energy efficiency, and latency, but also to support massive uplink traffic in order to provide occasional data transfer and Internet access for the massive number of spatially-spread  connected things.

To characterize the IoT performance, spatiotemporal mathematical models are required to simultaneously account for the massive spatial existence of the IoT devices as well as the traffic requirement per device. Due to the shared nature of the wireless channel, mutual interference is generated among the IoT devices with non-empty queues that operate on the same channel. The IoT devices are foreseen to exist in massive numbers and to have sporadic traffic patterns.  Hence, the mutual uplink interference between the IoT devices will be controlled via distributed transmission schemes rather than via centralized base station  (BS) scheduling~\cite{ NB-IOT, our_magazine}. Consequently, the queue and protocol states impose a fundamental impact on the aggregate interference generated in IoT networks, in which the magnitude of mutual interference between the IoT devices depends on their relative locations and the physical layer attributes. 

Characterizing the performance of wireless networks with explicit queues interactions among the nodes is a classical research problem in queueing theory ~\cite{21216, 70180, interacting_queues, 212177}.  However, the stand-alone queueing models in~\cite{21216, 70180, interacting_queues, 212177} adopt simple {\em collision model} to capture the interactions among the queues, which fails to account for interference based interactions that differ according to the distances and channel gains between the devices. Stochastic geometry is a solid mathematical tool to account for mutual interference between devices in large-scale networks~\cite{martin_book, survey_h, tutorial}. However, stochastic geometry based models are usually traffic agnostic and assume backlogged network with saturated queues. Stand alone stochastic geometry or queueing theory models can only obtain loose pessimistic bounds on the performance due to the massive numbers of IoT devices. Recently, \cite{queue1} integrates stochastic geometry and queuing theory to study sufficient and necessary conditions for queue stability in a network with spatially spread interacting queues. However, the model in \cite{queue1} is well suited for ad hoc networks with constant link distances and only derives stability conditions. 

This paper develops a novel spatiotemporal mathematical model for IoT-enabled by cellular networks.\footnote{{This paper is presented in part in \cite{Garbieh}.}} From the spatial domain perspective, stochastic geometry is used to model both of the cellular network BSs as well as the IoT devices using independent Poisson point processes (PPPs). In addition to simplifying the analysis, there are several empirical foundations that validate the PPP model for cellular networks~\cite{andrews2011tractable, martin_ppp,marco_fitting, queue1, survey_h, tutorial}. From the temporal domain perspective, two-dimensional discrete time Markov chain (DTMC) is employed to track the time evolution of the queue and the transmission protocol states of the IoT devices. The proposed system model abstracts the IoT-enabled cellular network to a network of spatially interacting queues, where the interactions reside in the mutual interference between the devices. 

\vspace{-3mm}
\subsection{Methodology, Contribution, and Organization}
While the queue arrival process is a function of the underlying IoT application only, the queue departure (i.e., service) process depends on the transmission signal-to-interference-plus-noise ratio (SINR) across time slots. We first show that the SINR involves a negligible temporal correlation across time slots due to the interference temporal correlation. Therefore, we drop the time index and assume independent SINR at each time slot. Then, we employ a two-dimensional Geo/PH/1 DTMC for each IoT device, where Geo stands for geometric inter-arrival process and PH stands for the Phase-type departure process \cite{MAM, alfa_DTMC}. The Geo/PH/1 queueing model is particularly chosen for its simplified memoryless inter-arrival process and its general departure process that can account for the interference based interactions between the IoT queues. To this end, we study three different transmission protocols that are defined by cellular systems for uplink random channel access~\cite{sesia2009lte}, namely, the baseline scheme, the power-ramping scheme, and backoff scheme. For each protocol, we derive sufficient conditions and necessary conditions for network stability, in which the stability is defined according to Loynes theorem \cite{stability}. That is, the network is considered stable if the average arrival rate is less than the average service rate for a randomly selected queue in the network.  The scalability and stability tradeoff in IoT cellular networks is then determined by plotting the Pareto-frontier of the stability region defined by the arrival process, the intensity of IoT devices, and the SINR detection threshold. The Pareto-frontier of the stability region shows the spatial traffic intensity limit (i.e., the maximum limit of IoT density for a given traffic requirement) that a cellular network can support, beyond which the queues become unstable and packets are lost with probability one.  For stable networks, we obtain the average queue size, the average packet delay, and the average service time. On the other hand, for unstable networks,  we only calculate the average service time. The contributions of this paper can be summarized by the following:
\begin{itemize}
\item Develop a novel spatiotemporal mathematical paradigm for IoT enabled cellular networks.
\item Integrate two dimensional Geo/PH/1 DTMC within stochastic geometry framework to account for interference based queue interactions in large-scale IoT network.
\item Derive sufficient and necessary conditions for IoT network stability. 
\item Characterize the scalability/stability tradeoff by showing the maximum spatial traffic intensity that a cellular network can accommodate. 
\item Assess and compare  three different transmission schemes in terms of transmission success probability, average packet delay, and average queue size.
\end{itemize}

{The rest of the paper is organized as follows. Section \ref{System Model} presents the system model and the assumptions. The outline of the analytical framework is established in Section \ref{framework}. Section \ref{Performance Analysis} characterizes the different transmission schemes. Section \ref{sec:Results} provides the numerical and simulation results. Finally, Section  \ref{Conclusions} concludes the paper.}


\vspace{-1mm}
\section{System Model and Assumptions}\label{System Model}
\vspace{-2mm}
\subsection{Spatial \& Physical Layer Parameters}
\vspace{-2mm}

We consider an IoT-enabled cellular network in which the cellular BSs and the IoT devices are spatially distributed in $\mathbb{R}^2$ according to two independent homogeneous PPPs, denoted as $\Psi$ and $\Phi$, with intensities $\lambda$ and $\mathcal{U}$, respectively.   Without loss of generality, each of the devices is assumed to communicate its packets via the nearest BS. The average number of devices connected to each BS is $\alpha = \frac{\mathcal{U}}{\lambda} $.
 
A power-law path-loss model is considered where the signal power decays at a rate $r^{-\eta}$ with the propagation distance $r$, where $\eta>2$ is the path-loss exponent.   In addition to the path-loss attenuation, Rayleigh fading multipath environment is  assumed where the intended and interfering channel power gains ($h$ and $g$) are exponentially distributed with unity mean. All channel gains are assumed to be independent of each other, independent of the spatial locations, and are identically distributed (i.i.d). The IoT devices use full path-loss inversion power control with a threshold $\rho$. That is, each device controls its transmit power such that the average signal power received at the serving BS is equal to a predefined value $\rho$, which is assumed to be the same for all the BSs. It is assumed that the BSs are dense enough such that each of the devices can invert its path loss towards the closest BS almost surely, and hence the maximum transmit power of the IoT devices is not a binding constraint for packet transmission. Since the main focus of this paper is to integrate stochastic geometry and DTMCs to study scalability and stability of IoT enabled cellular networks, we particularly select power-law path-loss, Rayleigh fading, and full-path-loss inversion for mathematical simplicity. Future extensions of this work may include multi-slope path loss models~\cite{Renzo_Intensity_Matching, Andrews_densification}, advanced fading models~\cite{MSC}, multi-antenna~\cite{Renzo_TCOM_Multi_antenna}, and/or fractional channel inversion power control~\cite{Renzo_TCOM_Multi_antenna, uplink2_jeff}. 

\vspace{-5mm}
\subsection{MAC Layer Parameters}
\vspace{-3mm}
This paper considers a synchronous time slotted cellular system and geometric packet inter-arrival process with parameter $a \in [0,1]$ (packet/slot) at each IoT device.  We assume that the duration of each time slot is small enough such that a single packet arrival and/or departure can take place per time slot.  The arrived packets at each IoT device are stored in a buffer (i.e., queue) with infinite length until successful transmission.\footnote{ The infinite buffer is assumed for mathematical convenience. In Section~\ref{sec:Results}, it is shown that small buffer sizes are sufficient as long as the network is stable. Otherwise, the buffers of the devices will overflow irrespective of their sizes.} The BSs are unaware about the queue status of the devices, and hence, no scheduling decision for the packets is taken at the BS side. Therefore, we assume that each device with non-empty queue directly transmits backlogged  packets via a First Come First Served (FCFS) service discipline, which complies with the narrow-band IoT (NB-IoT) operation defined by the 3GPP standard~\cite{NB-IOT}.   

A two-dimensional Geo/PH/1 Markov chain is employed to model each IoT device, in which the levels indicate the number of packets in the queue and the phases indicate the transmission protocol states.  Let $p_{m}$ be the probability of packet transmission success when the IoT device operate at transmission phase $m$ and let $(\bar{\cdot}) = (1 - \cdot)$ denote the probability complement operator. Then, the considered transmission protocols are shown in Figs.~\ref{fig_Markov_baseline}, \ref{fig_Markov_power_ramping}, and \ref{fig_Markov_backoff}, which are described below:

  \subsubsection{\textbf{Baseline scheme}} 
\begin{figure}[t!]
        \centering
    \includegraphics[width=3.5in]{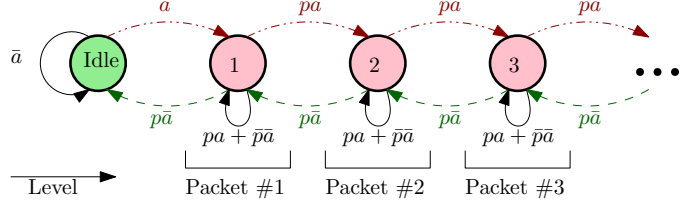}
\vspace{-5mm}
    \caption{ Queue aware schematic diagram for the baseline scheme. The green color indicates empty queue and hence idle state (not transmitting) and the red color indicates non-empty queue with transmission state.}
\label{fig_Markov_baseline}
\end{figure}

devices with non-empty queues keep transmitting backlogged packets in each time slot with the same power control threshold $\rho$.  A queue aware schematic diagram for the DTMC of the baseline scheme is shown in Fig.~\ref{fig_Markov_baseline}, where each level represents the number of packets in the queue. Note that the baseline scheme has a single protocol state for packet transmission, and hence, the phase subscript for $p_{1}$ is dropped and the scheme is represented by the single-dimension Markov chain shown in Fig.~\ref{fig_Markov_baseline}.


  \subsubsection{\textbf{Power ramping scheme}} 
\begin{figure}[t!]
        \begin{center}
	  \subfigure[Power-ramping Scheme.]{\label{fig_Markov_power_ramping}\includegraphics[width=3.in ]{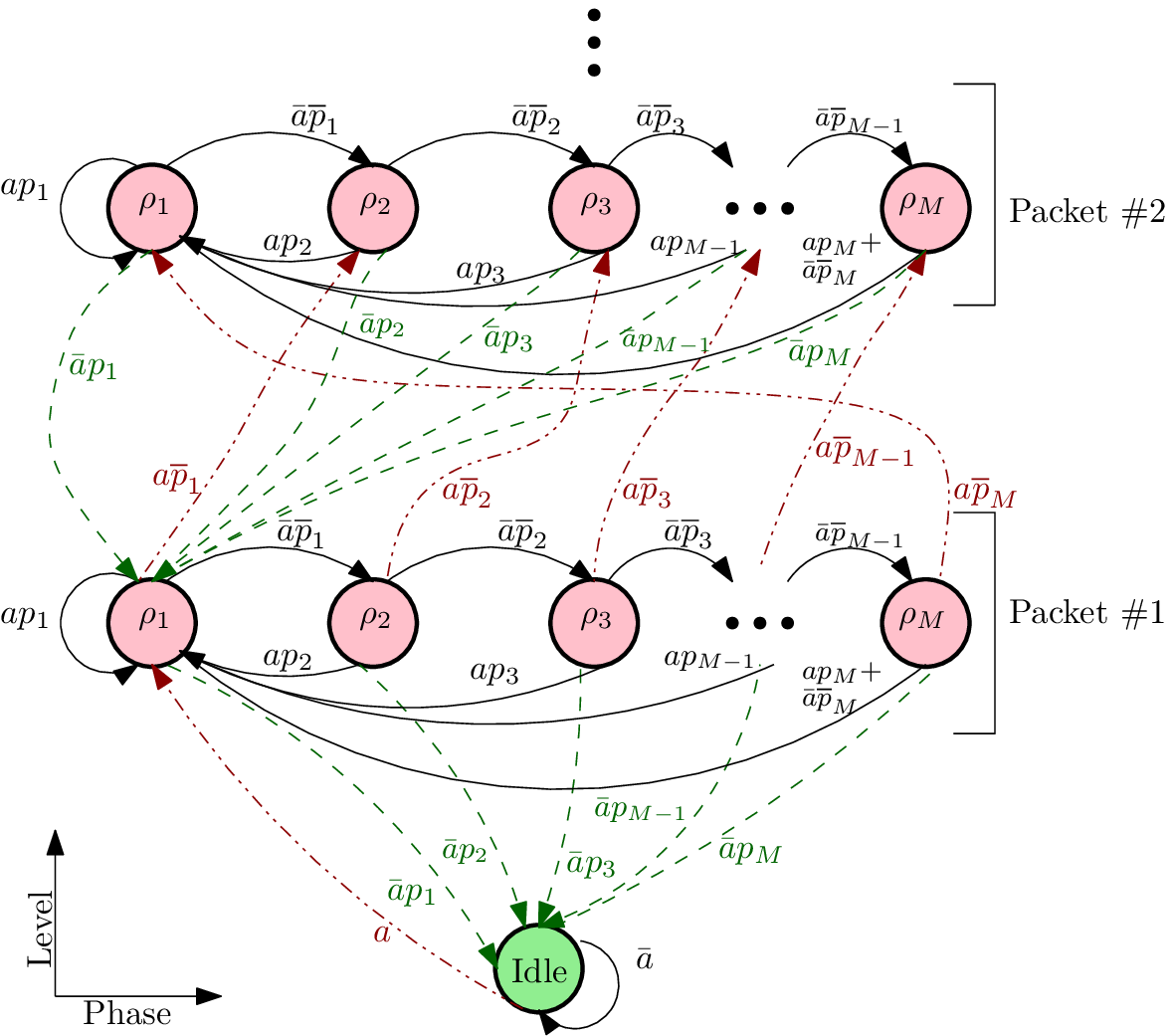}}
	  \subfigure[Backoff Scheme.]{\label{fig_Markov_backoff}
	  \includegraphics[width=3. in]{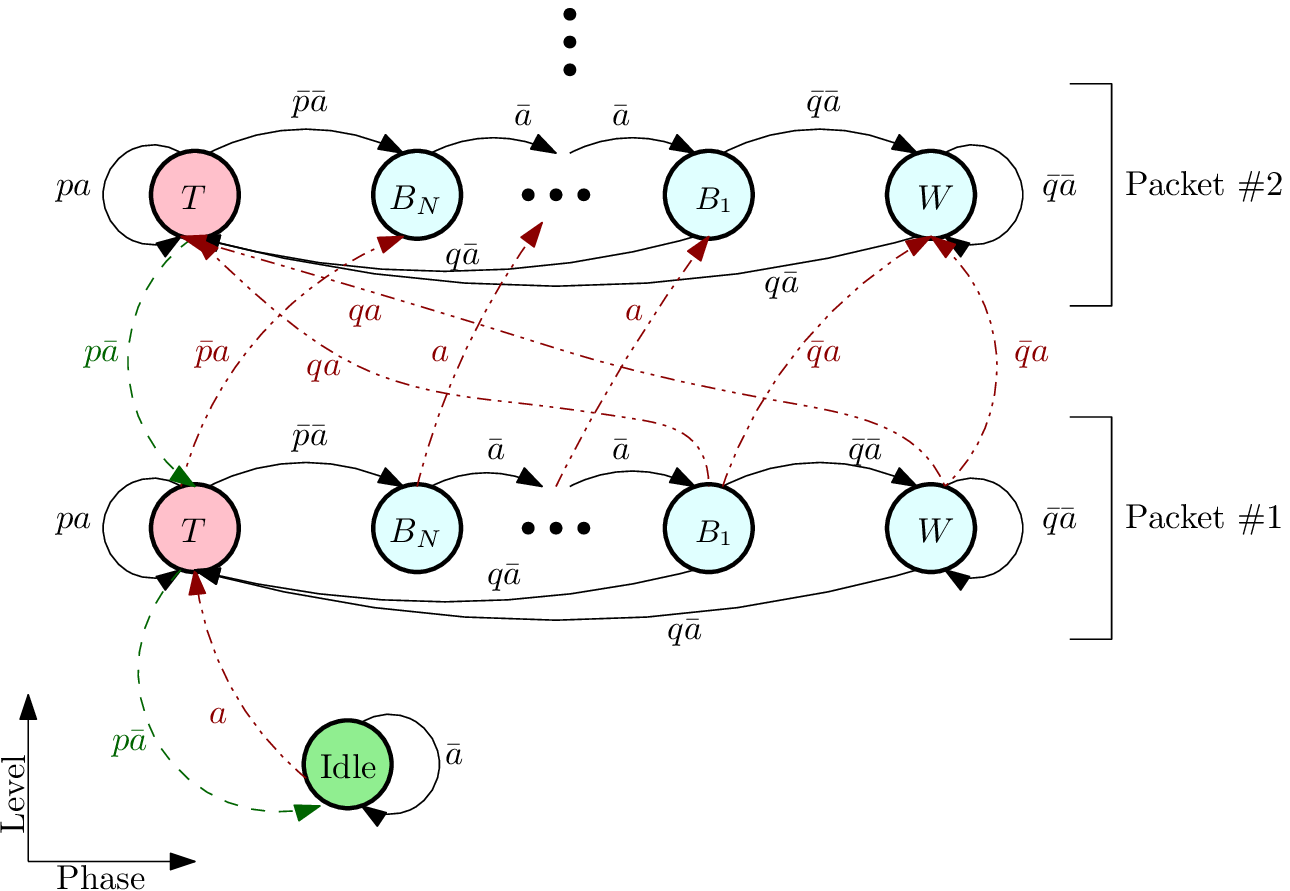}}
	\end{center}
	\vspace{-5mm}
	\caption{Queue and state aware DTMC for (a) the power-ramping scheme and (b) The backoff Scheme. The green color indicates empty queue and hence idle state (not transmitting), the red color indicates non-empty queue with transmission state, and cyan color indicates non-empty queue with backoff state.}\label{fig_Markov}
\end{figure}

\!\! devices with non-empty queues keep transmitting backlogged packets in each time slot, but with increasing power control threshold after each unsuccessful transmission until the maximum allowable threshold $\rho_M$ is reached. Let $\rho_m$ be the used power control threshold at the $m^{th}$ access attempt, then the power-ramping strategy enforces $\rho_1\! <\! \rho_2\!< \!\!\cdots\!\! <\! \rho_m < \!\! \cdots\!\! < \!\! \rho_M $ to prioritize delayed packets. Upon transmission success with non-empty buffer and/or the maximum number of retransmissions $M$ is reached, the device repeats the same strategy starting from the initial power control threshold $\rho_1$ to relief interference.  A queue and transmission state aware DTMC schematic diagram for the power-ramping scheme is shown in Fig.~\ref{fig_Markov_power_ramping}, where $p_{m}$ is the packet transmission success probability given that the device is using the power control threshold $\rho_m$. 

%

  \subsubsection{{\textbf{Backoff scheme}}} devices with non-empty queues attempt to transmit their packets, however, they defer their attempts upon transmission failure to alleviate congestion on the uplink channel. The backoff procedure contains deterministic backoff state for $N$ time slots followed by a probabilistic backoff state with probability $1-q$. The selected backoff scheme is general to capture deterministic backoff only by setting $q=1$, random backoff only by setting $N=0$, and generic combinations of both deterministic and random backoff states by setting $N\geq1$ and $q\leq1$. A queue and state aware DTMC schematic diagram for the device in the backoff scheme is shown in Fig.~\ref{fig_Markov_backoff}. Although there are multiple phases in the backoff scheme, there is only one transmission phase where the packet can be successfully transmitted. Hence, the phase subscript for the success probability in the transmission phase $p_{1} = p $ is dropped. 
%

 It is assumed that all IoT devices use one of the aforementioned transmission schemes to access the channel. Furthermore, the employed transmission scheme is used along with pseudo codes to alleviate congestion. For instance, the LTE defines prime-length Zadoff-Chu (ZC) sequences for random access channel (RACH)~\cite{sesia2009lte}. Without loss of generality, we assume that all BSs have the same number of pseudo codes and that different codes are orthogonal. 
 It is also assumed that each device randomly and independently selects one of the available codes at each transmission attempt. Hence,  the average number of devices that may use the same pseudo code per BS is $\tilde{\alpha} = \frac{\mathcal{U}}{\lambda n_Z} $, where $n_Z$ is the number of pseudo codes per BS.  Note that $\tilde{\alpha}$, measured in device/BS/code, is crucial network parameter that is  used later to study the scalability of cellular networks.



\vspace{-2mm}
\section{Outline of the Analytical Framework}  \label{framework}
\vspace{-2mm}
We employ the matrix-analytic-method (MAM) to analyze the Geo/PH/1 queueing model representing each IoT device~\cite{MAM, alfa_DTMC}, and stochastic geometry to analyze the interference based interactions among the queues. Before delving into the analysis details for each transmission scheme, this section describes the general analytical framework used in the analysis.
\vspace{-5mm}
\subsection{Queueing Analysis} \label{queue_analysis}
\vspace{-2mm}
 From the queueing perspective, the PH type distribution tracks the phases that the queue may experience until the packet is successfully transmitted. Let $n$ be the number of transient phases. Following \cite{alfa_DTMC}, the PH type distribution is defined by the tuple ($\boldsymbol{\beta},\mathbf{S}$), where $\mathbf{S} \in \mathbb{R}^{n\times n}$ is the transient sub-stochastic matrix and $\boldsymbol{\beta} \in \mathbb{R}^{1\times n}$ is the initialization row-vector for the transient states.  Particularly, the PH type distribution models the service procedure with an absorbing Markov chain with the following transition matrix 
\vspace{-1mm}
\begin{align}
\small
\mathbf{\mathcal{P}}=\begin{bmatrix}
1 & 0   \\ 
\mathbf{s} & \mathbf{S} \\ 
\end{bmatrix},
\label{absorb}
\normalsize
\end{align}
where $\mathbf{s} \in \mathbb{R}^{n \times 1}$ is given by $\mathbf{s} =\mathbf{e}-\mathbf{S} \times \mathbf{e}$ and $\mathbf{e}$ is a column vector of ones with the proper length. From \eqref{absorb}, it is clear that $\mathbf{S}$ captures the transition probabilities between phases until absorption and $\mathbf{s}$ captures the probability of absorption from each phase, in which absorption corresponds to successful packet transmission through the channel. Exploiting  the MAM with PH type service along with the packet-by-packet generation and service assumption, each IoT device employing any of the aforementioned transmission schemes can be represented by a Quasi-Birth-Death (QBD) queueing model with the following general probability transition matrix
\begin{align}
\small
\mathbf{P}=\begin{bmatrix}
B & \bold{C} &  &  &  & \\ 
\bold{E} & \bold{A}_1 & \bold{A}_0 &  &  &\\ 
  &\bold{A}_2  & \bold{A}_1 & \bold{A}_0 &   &\\ 
  &  & \bold{A}_2 & \bold{A}_1 & \bold{A}_0&  &\\ 
  &  &     &\ddots  &\ddots  &\ddots   
\end{bmatrix},
\normalsize
\label{eq:trans_matrix1}
\end{align}
where $B \in\mathbb{R} ,\bold{C} \in \mathbb{R}^{1\times n}, \bold{E} \in \mathbb{R}^{n \times1}, \bold{A}_0 \in \mathbb{R}^{n \times n}, \bold{A}_1 \in \mathbb{R}^{n \times n}$, and $\bold{A}_2 \in \mathbb{R}^{n \times n}$ are the sub-stochastic matrices that capture the transitions between the queue levels. Particularly, the sub-matrix $\bold{A}_0=a \mathbf{S}$ captures the event where a packet arrives while being in the transient service state, which increases the number of packets in the queue by one. The sub-matrix $\bold{A}_2 =\bar{a} \mathbf{s}\boldsymbol{\beta}$ captures the event where a service completion occurs and no packet arrives, which decreases the number of packets in the queue by one.  The sub-matrix $\bold{A}_1 =a \mathbf{s}\boldsymbol{\beta} +\bar{a} \mathbf{S}$ captures the events where a service completion occurs and a packet arrives or no packet arrives while being in transient state, which leaves the number of packets in the queue unchanged.  The transitions captured by $\bold{A}_2$, $\bold{A}_1$, and $\bold{A}_0$ are visualized in Figs.~\ref{fig_Markov_baseline}, \ref{fig_Markov_power_ramping}, and \ref{fig_Markov_backoff} by green, red, and black arrows, respectively. Boundary vectors $B=\bar{a},\bold{C}=a \boldsymbol{\beta}$, and $\bold{E}  = \bar{a} \mathbf{s}$ capture the transitions from idle-to-idle, idle-to-level (i=1), and from level (i=1)-to-idle, respectively. Note that $B,\bold{C}$, and $\bold{E} $ have different sizes than $\bold{A}_2$, $\bold{A}_1$, and $\bold{A}_0$ due to the different phase structure between the idle state and the other levels $(i \geq 1)$.  It is worth noting that the vector $\boldsymbol{\beta}$ appears in $\bold{C}, \bold{A}_1,$ and $\bold{A}_2$ to initialize the transient service states for a new packet. 

\begin{figure}[t!]
        \begin{center}
	  \subfigure[Power-ramping Scheme.]{\label{fig_Markov_power_ramping_nonstable}\includegraphics[width=3in]{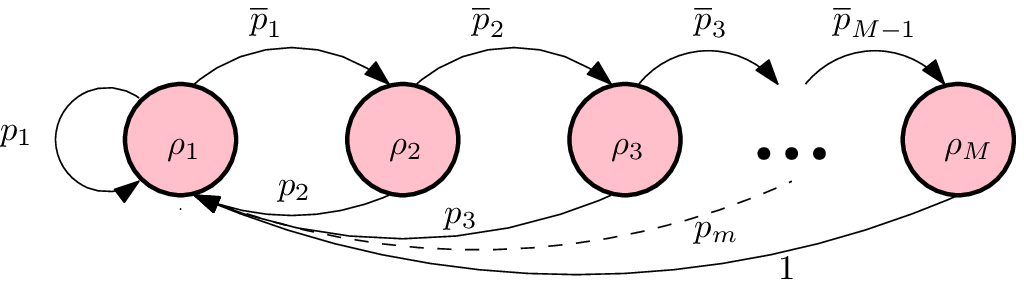}}
	  \subfigure[Backoff Scheme.]{\label{fig_Back_off_nonstable_nonstable}\includegraphics[width=3in]{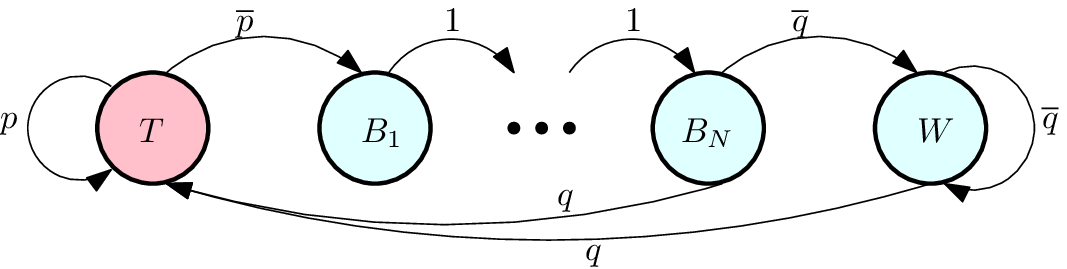}}
	\end{center}
	\vspace{-5mm}
    \caption{State aware DTMC for unstable network operation.}
\end{figure}

Let $\bold{A}=\bold{A}_0+\bold{A}_1+\bold{A}_2$, and let the vector $\boldsymbol{\pi}$ be the unique solution of $\boldsymbol{\pi}  \bold{A}=\boldsymbol{\pi}  $ with the normalization condition $ \boldsymbol{\pi}  \mathbf{e}=1$. 
Then, the DTMC in \eqref{eq:trans_matrix1} is stable if and only if $\boldsymbol{\pi} \bold{A}_2  \mathbf{e}>\boldsymbol{\pi}  \bold{A}_0  \mathbf{e}$,
 which implies that the departure rate is higher that the arrival rate \cite{alfa_DTMC,MAM}.
\noindent For stable systems, the steady state solution can be obtained by solving 
\vspace{-2mm}
\small
\begin{align}
\mathbf{x}\mathbf{P}=\mathbf{x}\text{,} \quad \mathbf{x} \mathbf{e} =1,
\label{eq:distribution1}
\end{align}
\normalsize
\vspace{-2mm}
where $\mathbf{x}=[x_\circ, x_{1,1} x_{1,2}, \dots, x_{i,j-1}, x_{i,j}, x_{i,j+1} \dots]$ is the row vector that contains the steady-state probabilities, in which $x_\circ$ denotes the idle state probability and $x_{i,j}$ is the probability of being in level $i$ and phase $j$. Note that $\mathbf{x}$ is also denoted as the steady state distribution of the queue. When the stability condition ($\boldsymbol{\pi} \bold{A}_2  \mathbf{e}>\boldsymbol{\pi}  \bold{A}_0  \mathbf{e}$) is not satisfied, the queue becomes unstable and the queue length grows indefinitely and the system in \eqref{eq:distribution1} with the transissiom matrix in \eqref{eq:trans_matrix1} cannot be analyzed \cite{MAM, alfa_DTMC}. Consequently, we abstract the queue size for unstable queues and look into the marginal phase distribution only $\bold{\Pi}$ as shown in Figs. \ref{fig_Markov_power_ramping_nonstable} and \ref{fig_Back_off_nonstable_nonstable}, which is obtained by solving the following system
\vspace{-2mm}
\small
\begin{align}
\mathbf{\bold{\Pi}}\mathbf{P}=\mathbf{\bold{\Pi}}\text{,} \quad \mathbf{\bold{\Pi}} \mathbf{e} =1\text{,} \quad \text{where} \quad \mathbf{P}=   \mathbf{s}  \boldsymbol{\beta} + \mathbf{S} ,
\label{eq:distribution2}
\end{align}
\normalsize
\vspace{-2mm}
such that $\boldsymbol{\Pi}$ contains the probabilities for the states shown in Figs. \ref{fig_Markov_power_ramping_nonstable} and \ref{fig_Back_off_nonstable_nonstable}. Note that these systems are finite and ergodic, and hence, are always stable.

To analyze the queue stability and solve the above queueing systems, the PH type distribution parameters ($\boldsymbol{\beta},\mathbf{S}$) should be determined. As discussed earlier and shown in Figs.~\ref{fig_Markov_baseline}, \ref{fig_Markov_power_ramping}, and \ref{fig_Markov_backoff}, the transmission protocol restarts its phases after each successful transmission. Hence, $\boldsymbol{\beta}\!\! =\!\!\{1,\boldsymbol{0}\}$ where $\boldsymbol{0}$ is a row vector of zeros with the proper length. This makes $\mathbf{S}$ the only missing component to solve the presented queueing model. Figs.~\ref{fig_Markov_baseline} and \ref{fig_Markov}  show that $\mathbf{S}$ involve computing the transmission probabilities $p$ and $p_{m}$, which are discussed in the next section. 

 

\vspace{-3mm}
\subsection{Stochastic Geometry Analysis for Service Probability} \label{SG_analysis}
\vspace{-1mm}
A packet sent by an IoT device (i.e., queue) is successfully decoded at the serving BS (i.e., server) if the SINR  at the BS exceeds a certain threshold $\theta$. Mathematically, the transmission success probabilities for the baseline and backoff schemes are expressed as
\vspace{-.1mm}
\small
\begin{align}
p&=\mathbb{P}\Bigg\{ \frac{\rho h_\circ }{  \underset{\triangleq \mathcal{I}}{\underbrace{ \sum\limits_{y_i \in \tilde{\Phi}} P_i g_i \left\| y_i - z_\circ \right\|^{-\eta}}}+ \sigma^2}> \theta \Bigg\}  = \exp\left\{- \frac{\sigma^2 \theta}{\rho} \right\}  \mathcal{L}_{\mathcal{I}} \left(\frac{\theta}{\rho} \right),
\label{sinr}
\end{align}
\normalsize

\noindent where $\left\| \cdot \right\|$ is the Euclidean norm, $\mathcal{L}_{\mathcal{I}}(\cdot)$ denotes the Laplace transform (LT) of the PDF of the aggregate interference $\mathcal{I}$, $h_\circ$ is the intended channel gain,  $\tilde{\Phi}$ is the set containing the locations of interfering devices, $P_i$, $g_i$, and $y_i \in \mathbb{R}^2$ are, respectively, the transmit power, the channel power gain, and  location of the $i^{th}$ interfering device, $z_\circ \in \Psi$ is the location of the serving BS of the test device, $\sigma^2$ is the noise power, and $\theta$ is the SINR detection threshold. Note that  \eqref{sinr} follows from the channel inversion power control, the exponential distribution of $h_\circ$, and the definition of the Laplace transform~\cite{elsawy2014stochastic}.  Similarly, the success probability for the power-ramping scheme for the test device when using power control threshold $\rho_m$ is given by

\vspace{-5mm}
\small
\begin{align}
p_{m}&=\mathbb{P}\Bigg\{ \frac{\rho_m h_\circ }{  \sum\limits_{k=1}^M \underset{\triangleq \mathcal{I}_k}{\underbrace{ \sum\limits_{y_i \in \tilde{\Phi}_{k}} P_{i,k} g_i \left\| y_i - z_\circ\right\|^{-\eta}}}+ \sigma^2}> \theta \Bigg\}{=} \exp\left\{- \frac{\sigma^2 \theta}{\rho} \right\}  \prod_{k=1}^M \mathcal{L}_{\mathcal{I}_k} \left(\frac{\theta}{\rho_m} \right),
\label{sinr_ramp}
\end{align}
\normalsize
where $\tilde{\Phi}_{k}$ is the set containing the locations of interfering devices operating with power control $\rho_k$, $P_{i,k}$ is the transmit power of the $i^{th}$ device in $\tilde{\Phi}_{k}$, $\mathcal{I}_k$ is the aggregate interference power from devices in $\tilde{\Phi}_{k}$. The transmit powers $P_i$ and $P_{i,k}$ in \eqref{sinr} and \eqref{sinr_ramp} are random variables that depend on the distance between the i$^{th}$ interfering IoT device and its own serving BS. The expression in \eqref{sinr_ramp} follows from the employed channel inversion power control, the exponential distribution of $h_\circ$, the definition of the Laplace transform, and the independence between $\tilde{\Phi}_k$, $\forall k$ at a given time slot. Equations \eqref{sinr} and \eqref{sinr_ramp} capture the interactions between the queues in the aggregate interference term. The equations also involve the effect of propagation, fading, and power control on the transmission success probabilities.  The interference terms in \eqref{sinr} and \eqref{sinr_ramp} are functions of the points processes $\tilde{\Phi}$ and $\tilde{\Phi}_{k}$, respectively. Due to the random pseudo code selection and the independent thinning property of the PPP, both $\tilde{\Phi}$ and $\tilde{\Phi}_{k}$ are PPPs at any time slot. However, there may exist common interfering devices seen by the same BS from one time slot to another, which introduces temporal correlation for the success probability. Such temporal correlation involves memory to the queues and highly complicates the queueing analysis. Furthermore, the transmission powers of nearby devices are spatially correlated due to the employed channel inversion power control. Such temporal and spatial correlations impede the model tractability. Fortunately, the aforementioned temporal and spatial correlations have a negligible impact on the performance of the employed system model, which is verified by numerical results and simulations. Formal statements and comments on the involved  approximations are given in the sequel. 

\vspace{-7mm}
\begin{approximation} The spatial correlation among the transmission powers of the interfering IoT devices is ignored.
\end{approximation}
\vspace{-5mm}
\begin{remark}
 The spatial correlation among the transmission powers of interfering devices is a common feature uplink systems~\cite{Andrews_TWC_Uplink, Renzo_TCOM_Multi_antenna, elsawy2014stochastic, Kim_Uplink, uplink2_jeff, marco_uplink}, which is always ignored to maintain mathematical tractability. It is shown that such spatial correlation of the interfering powers is weak and have a negligible effect on the success probability~\cite{Andrews_TWC_Uplink, Renzo_TCOM_Multi_antenna, elsawy2014stochastic, Kim_Uplink, uplink2_jeff, marco_uplink}. In Section~\ref{sec:Results}, we verify the accuracy of such approximation within our mathematical model. It is worth noting that the devices in our system model transmit according to a distributed transmission  protocol (i.e., baseline, power ramping, and backoff), which is different from the centralized BS scheduling in~\cite{Andrews_TWC_Uplink, Renzo_TCOM_Multi_antenna, elsawy2014stochastic, Kim_Uplink, uplink2_jeff, marco_uplink}. Consequently, the uplink transmission in the considered IoT scenario experience intra-cell and inter-cell interference.
 \end{remark}
\vspace{-5mm}


 \begin{figure}[t!]
        \centering
    \includegraphics [width=2.3 in]{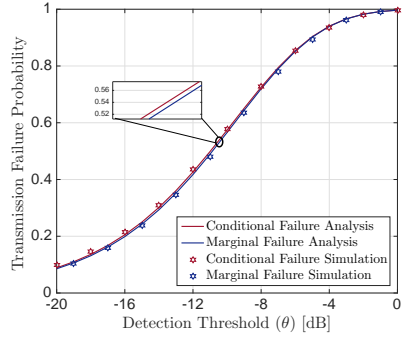}
    \vspace{-5mm}
    \caption{The effect of temporal interference correlations on the transmission failure probability for $\eta=4$, $\tilde{\alpha}=4$, $\rho = -90$ dBm, $\sigma^2 = -90$ dBm, and $\eta=4$  when the probability of having common interfering sources is $\frac{1}{2}$ across time slots.}
\label{fig:correlation}
\vspace{-2mm}
\end{figure}


\begin{approximation} Under the employed full channel inversion power control, the temporal interference correlation has a negligible effect on the transmission success probability. Hence, we ignore the temporal SINR correlation and assume that the test IoT device sees almost independent $\tilde{\Phi}$  and $\tilde{\Phi}_{n}$ in each time slot.\footnote{It is important to note that Approximation 2  is only accurate for the full channel inversion power control. For constant power uplink or downlink transmissions, such accuracy may not hold. }
\end{approximation}
\vspace{-5mm}
\begin{remark}
 Let $\mathbb{P}\left\{\rm{SINR_{t_2}}\! <\! \theta \vert \rm{SINR_{t_1}}\! <\! \theta \right\}$  be the conditional  failure probability (i.e., the complement of the success probability),  where ${\rm SINR_{t_i}}$  is the SINR at the $i^{th}$ time slot.  Appendix A derives a closed form expression for $\mathbb{P}\left\{\rm{SINR_{t_2}}\! <\! \theta \vert \rm{SINR_{t_1}}\! <\! \theta \right\}$ in \eqref{eq:corr1}. Fig.~\ref{fig:correlation}  compares the conditional success probability derived in \eqref{eq:corr1} to  the marginal success probability derived in \eqref{eq:Out1} for an aggressive scenario where all IoT devices are active and only two pseudo codes are available. The negligible effect of the temporal correlation shown in  Fig.~\ref{fig:correlation}, even when the test BS sees $50\%$ common interfering devices from one time slot to another, strongly supports Assumption 1. The system model used in this manuscript impose far less probability of having common interferers across time slots due to the availability of multiple pseudo codes (typically 64 code~\cite{sesia2009lte}) and the probability of having idle/backoff states, which further diminishes the effect of interference temporal correlation~\cite{Haenggi_TWC_Managing}. 
\end{remark}

\noindent Based on Approximation 2, it is reasonable to assume that all devices experience i.i.d. steady state queue distributions (i.e., $\mathbf{x}$) irrespective of their locations. Let $x_\circ$ denote the probability of the idle state and let $\boldsymbol{{\Pi}}$ be the steady state marginal distribution for all protocol states such that $\Pi_j=\sum_{i=1}^\infty x_{i,j}$. Then the intensity for $\tilde{\Phi}$ in the baseline and backoff schemes are $\tilde{\mathcal{U}} = \frac{(1-x_\circ) \mathcal{U}}{n_Z}$ and $\tilde{\mathcal{U}} = \frac{\Pi_1 \mathcal{U}}{n_Z}$, respectively. Similarly, the intensity of  $\tilde{\Phi}_{k}$ in the power-ramping scheme is $\tilde{\mathcal{U}}_k = \frac{\Pi_k \mathcal{U}}{n_Z}$. Note that the aforementioned intensities of interfering IoT devices are functions of the power control, traffic profile, transmission  protocol, relative devices to BS densities, and aggregate interference, which is different from the load-aware intensity adjustments that are functions of the relative devices to BS densities only \cite{Andrews_offloading, Renzo_Intensity_Matching , offload_elsawy, uplink_alamouri, Andrews_Downlink_Distribution }. As will be shown later, the interfering devices intensity in our mathematical model depends on the steady state marginal distribution $\boldsymbol{{\Pi}}$ (or equivalently the steady state probability vector $\mathbf{x}$), which imposes the causality problem discussed in the next section.

\vspace{-4mm}
\subsection{Causality problem and the iterative solution} \label{causal}
\vspace{-2mm}
The queueing and stochastic geometry analysis discussed in Sections \ref{queue_analysis} and \ref{SG_analysis} are interdependent, which impose a causality problem.  Particularly, the matrix $\mathbf{S}$ that is required to construct and solve the queuing model in Section \ref{queue_analysis} depends on the transmission success probabilities in \eqref{sinr} and \eqref{sinr_ramp}  of Section \ref{SG_analysis}. The converse is also true, the point processes $\tilde{\Phi}$  and $\tilde{\Phi}_{n}$ in \eqref{sinr} and \eqref{sinr_ramp} of  Section \ref{SG_analysis} are constructed based on the steady state solution of \eqref{eq:distribution1} in Section \ref{queue_analysis}. To solve this causality problem, we employ the iterative solution, shown in Algorithm 1, to find the steady state probabilities and the associated aggregate interference. 
\begin{figure*}
\begin{algorithm}[H]
\small
\textbf{Initialize} $x_\circ^{[0]}$ and $\boldsymbol{{\Pi}}^{[0]}$ such that $x_\circ+\boldsymbol{{\Pi}}^{[0]} \times \mathbf{e} = 1$.\\
\textbf{Set} i =1.\\
			1- Use $\boldsymbol{{\Pi}}^{[i-1]}$ to calculate the transmission success probabilities in \eqref{sinr} and \eqref{sinr_ramp} via stochastic geometry.\\
  				2- Use the transmission success probabilities from step 1 to construct the matrix $\textbf{S}$. \\
				3- Check the stability condition  ($\boldsymbol{\pi} \bold{A}_2  \mathbf{e}>\boldsymbol{\pi}  \bold{A}_0  \mathbf{e}$) then 	\\
\eIf{stable}{
   Use the MAM to solve the QBD system in \eqref{eq:distribution1} using $\textbf{P}$ in \eqref{eq:trans_matrix1} and obtain $\mathbf{x}$.\;
   }{
    Use the MAM to solve the system in \eqref{eq:distribution2} using $\textbf{P}$ in \eqref{eq:distribution2} and obtain $\mathbf{x}$. \;
  }				 
				4- Use the steady state distribution $\mathbf{x}_0^{[i]}$  to update $x_\circ$ and the marginal distribution $\boldsymbol{{\Pi}}^{[i]}$.  \\
				\textbf{Repeat} steps 1 to 4 until  { $ \max| \boldsymbol{\Pi}^{[i]} -  \boldsymbol{\Pi}^{[i-1]} |\leq \epsilon $  } 
  \textbf{Return} $\boldsymbol{\Pi} \leftarrow \boldsymbol{\Pi}^{[i]}$.		
\caption{General Framework for the Analysis}
\normalsize
\end{algorithm}
\vspace{-12mm}
 \hrulefill
\end{figure*}
The iterative algorithm in Algorithm 1 is equivalent to solving the system 
\vspace{-.1mm}
\small
\begin{align}
\mathbf{x}^{[i]}\mathbf{P}\!\!\left(\mathbf{x}^{[i-1]}\right)=\mathbf{x}^{[i]}  \quad \text{,} \quad {\mathbf{x}^{[i]}}^T \mathbf{e} =1,
\label{eq:iter_distribution}
\end{align}
\normalsize
where $\mathbf{P}\!\!\left(\mathbf{x}^{[i-1]}\right)$  is updated based on \eqref{sinr} and \eqref{sinr_ramp}. Due to the ergodicity of the used queueing models, the system in \eqref{eq:iter_distribution} has a unique solution $\mathbf{x}^{[i]}$ for each $\mathbf{x}^{[i-1]}$~\cite{alfa_DTMC}. Hence, the {Algorithm~1} converges by virtue of the fixed point theorem~\cite{convergence}. 

\vspace{-7mm}
\subsection{Performance Metrics}
\vspace{-1mm}
After determining the steady state solution $\mathbf{x}$ for the system via Algorithm 1, many performance metrics can be obtained. For instance, transmission success probability at steady state is 
\vspace{-3mm}
\small
\begin{equation}
p_m=\mathbb{P}\left\{ \Upsilon_m\!(\boldsymbol{\Pi}) > \theta \right\},
\vspace{-3mm}
\end{equation}
\normalsize

\noindent where  $\Upsilon_m\!(\boldsymbol{\Pi})$ is the SINR when using transmission phase $m$ and the activity of the nodes is determined by $\boldsymbol{\Pi}$,  in which $\boldsymbol{\Pi}$ is the steady state marginal distribution of the phases. Constructing an absorbing Markov chain for the phases the packet sees until successful transmission, the expected transmission delay can be estimated using \cite[eq(2.24)]{alfa_DTMC}      as: 
\normalsize
\vspace{-2mm}
{\begin{equation}\label{ave_retrans}
\mathcal{D}= \frac{1}{ \mathbb{E}[ p_m] }= \frac{1-x_\circ}{\sum_m \Pi_m p_m } ,
\vspace{-2mm}
\end{equation}}
\normalsize
\noindent where the expectation in the denumerator accounts for the different states that the IoT device may be in, $x_\circ$ is the probability of being idle, $\Pi = [\Pi_1, \Pi_2, .......]$ is the marginal distribution of the phases.

For stable systems, we can additionally find the average queue size, denoted as $\mathbb{E}[Q_L]$, and the average queueing delay, denoted as  $\mathbb{E}[W_q]$. Let $Q_L$ be  instantaneous queue size, then the average queue size in given by
\vspace{-2mm}
\small
 \begin{align} \label{ave_queue}
\mathbb{E}\left[Q_L\right] &= \sum_{n=1}^{\infty} n \mathbb{P} \left\{Q_L=n \right\} =\sum_{n=1}^{\infty} n \sum_{j} x_{n,j}.
\vspace{-2mm}
\end{align}
\normalsize

\noindent {The average waiting time for packet transmission accounts for the queueing delay only.} Let $W$ be the queueing delay (i.e., the number of time slots spent in the queue before the service starts) for a randomly selected packet, then the average queueing delay is given by \cite{alfa_DTMC} as 

\vspace{-2mm}
\small
\begin{equation} \label{wait_gen}
\mathbb{E}\left[W_q\right]=\sum_{n=1}^{\infty}n \mathbb{P}\left\{W=n \right\}, 
\end{equation}
where,
\vspace{-3mm}
\begin{align}
\mathbb{P} \left\{W=0 \right\} =x_\circ, \; \text{and} \quad \mathbb{P}\left\{W=j \right\} =\displaystyle{\sum\limits_{v=1}^{j} x_v B_j^{(v)}}, \; \text{with} \quad B^{(k)}_j  = \left\{\begin{matrix}
 \textbf{S}^{j-1} \textbf{s}  & k=1, j \geq 1  \\
(\textbf{s} \boldsymbol{\beta})^k  &   j=k, k\geq 1 \\
\textbf{S}B^{(k)}_{j-1}+\textbf{s} \boldsymbol{\beta}B^{(k-1)}_{j-1}  & k \geq j\geq 1 \\ 
\end{matrix}\right..
\end{align}
\normalsize


\noindent The average queueing plus transmission delay is then obtained as $\mathbb{E}\left[W_q\right] + \mathcal{D}$.

\vspace{-4mm}
\section{Performance Analysis}\label{Performance Analysis}

For the sake of organized presentation, we divide this section into three parts corresponding to each transmission scheme. For each transmission scheme we first analyze the necessary condition and sufficient condition for stability, then we conduct the analysis as discussed in Section~\ref{framework}. 
\vspace{-4mm}
\subsection{Baseline Scheme}
\vspace{-2mm}
We start by the baseline scheme, which is the simplest transmission scheme. Fig.~\ref{fig_Markov_baseline} shows that the baseline scheme has only one transmission phase. Hence, the transient state of \eqref{absorb} is just $\textbf{S}=\bar{p}$, where $p$ is the transmission success probability given by \eqref{sinr}. This reduces the Geo/PH/1 queueing model in \eqref{eq:trans_matrix1} to a simple Geo/Geo/1 queue with the following transmission matrix

\small
\begin{align}
\mathbf{P}=\begin{bmatrix}
\bar{a} & a &  &  &  & \\ 
\bar{a} p & \bar{a} \bar{p} + ap & a\bar{p} &  &  &\\ 
  &\bar{a} p & \bar{a} \bar{p} + ap & a\bar{p} &   &\\ 
  &  & \bar{a} p & \bar{a} \bar{p} + ap & a\bar{p} &  &\\ 
  &  &     &\ddots  &\ddots  &\ddots   
\end{bmatrix}.
\label{baseline_matrix}
\end{align}
\normalsize
Let $\mathbf{x}=[x_\circ,\ x_1,\ x_2,\ \hdots,\ ]$ be the stationary distribution, where $x_i$ represents the probability of having $i$ packets in the queue, then solving \eqref{eq:distribution1} with the transition matrix in \eqref{baseline_matrix} gives 
\begin{align}
\begin{matrix}
x_i=R^i \frac{x_\circ}{\bar{p}}, & \text{where} \quad R=\frac{a\bar{p}}{ \bar{a}p} & \text{and} \quad x_\circ=\frac{p-a}{p}.
\end{matrix}
\label{eq_base}
\end{align}
 Using stochastic geometry, the transmission success probability $p$ for the baseline scheme is characterized via the following lemma.

\begin{lemma} \label{lem:out_base}
The transmission success probability in the depicted IoT network with baseline transmission scheme is given by 
\small
\begin{align}\label{eq:Out1}
  {p}   & \;\;\; {\approx}  \;\;\;\frac {\exp\left\{- \frac{\sigma^2 \theta}{\rho} -  \frac{2 \theta \bar{x}_\circ \tilde{\alpha}}{(\eta-2) } \;{}_2F_1\left(1,1-2/\eta,2-2/\eta,-\theta\right) \right\}}{ \left(  1+  \frac{\theta \bar{x}_\circ \tilde{\alpha}}{(1+\theta) c}\right)^{c} } 
   \overset{(\eta=4)}{=} \;\;\; \frac{\exp\left\{- \frac{\sigma^2 \theta}{\rho} - \bar{x}_\circ \tilde{\alpha} \sqrt{\theta}\arctan\left({\sqrt{\theta}}\right)   \right\}}{ \left(  1+  \frac{\theta \bar{x}_\circ \tilde{\alpha}}{(1+\theta)c}\right)^{c} }
\end{align}
\normalsize
 where $c=3.575$, ${}_2 F_1(.)$ is the Gaussian hypergeometric function, and the approximation is due to the employed approximate PDF of the $\mathbb{R}^2$ PPP Voronoi cell area in \eqref{pdf_users} and Approximation 1.
\end{lemma}
\vspace{-5mm}
\begin{IEEEproof}
See Appendix \ref{sec:AppA}.
\end{IEEEproof}

The case of  $\eta=4$, which is a typical path loss exponent for urban outdoor environment, shown in \eqref{eq:Out1} is of interest because it simplifies the transmission success probability in terms of the elementary $\arctan(\cdot)$ instead of the Gaussian hypergeometric function. 

The interdependence between $x_\circ$ in \eqref{eq_base} and $p$ in \eqref{eq:Out1} clearly shows the causality problem discussed in Section~\ref{causal}. Although an iterative solution similar to Algorithm~1 is required to solve \eqref{eq_base} and \eqref{eq:Out1},  we can still obtain explicit conditions for network stability as shown in the following lemma and corollary.
\vspace{-3mm}  
\begin{lemma} \label{lem_stab_baseline}
For the baseline scheme in the depicted IoT network, the condition shown in \eqref{suf_stab_baeline} is sufficient for network stability and the condition shown in \eqref{necc_stab_baeline} is necessary for network stability.
\vspace{-3mm} 
\small 
\begin{align}\label{suf_stab_baeline}
\frac {\exp\left\{- \frac{\sigma^2 \theta}{\rho} -  \frac{2 \theta  \tilde{\alpha}}{(\eta-2) } \;{}_2F_1\left(1,1-2/\eta,2-2/\eta,-\theta\right) \right\}}{a \left(  1+  \frac{\theta}{(1+\theta)} \frac{  \tilde{\alpha}}{c}\right)^{c} } >1
 \end{align}
\vspace{-1mm}
\begin{align}\label{necc_stab_baeline}
 \frac {\exp\left\{- \frac{\sigma^2 \theta}{\rho} -  \frac{2 \theta  a\tilde{\alpha}}{(\eta-2) } \;{}_2F_1\left(1,1-2/\eta,2-2/\eta,-\theta\right) \right\}}{a \left(  1+  \frac{\theta}{(1+\theta)} \frac{ a \tilde{\alpha}}{c}\right)^{c} } >1
 \end{align} 
 \normalsize
 

\end{lemma}
\begin{IEEEproof}
The queue is stable if and only if the service probability is higher than the arrival probability $p>a$.  Substituting $p$ from \eqref{eq:Out1} with the maximum and minimum activity factors (i.e., $\bar{x}_\circ=1$ and $\bar{x}_\circ=a$), the sufficient and necessary inequalities are obtained, respectively. 
\end{IEEEproof}

From Lemma~\ref{lem_stab_baseline}, upper bounds on $\tilde{\alpha}$ can be obtained from the following corollary.

\begin{corollary} \label{col_stab_baseline}
Upper bounds for $\tilde{\alpha}$ device/BS/code, such that the necessary condition and sufficient conditions in \eqref{suf_stab_baeline} and  \eqref{necc_stab_baeline} hold, are respectively given in as 
\vspace{-1mm}
\small
\begin{align}\label{suf_stab_baeline1}
\tilde{\alpha} \leq \left(\frac{1}{\sqrt[c]{a}}-1\right) \frac{(\theta+1) c}{\theta} \quad \text{and} \quad \tilde{\alpha} \leq \left(\frac{1}{\sqrt[c]{a}}-1\right) \frac{(\theta+1) c}{ a \theta}.
\end{align}
\normalsize

\end{corollary}
\begin{IEEEproof}
 Taking the $\log(\cdot)$ for the both sides of \eqref{suf_stab_baeline} and \eqref{necc_stab_baeline}, it is noticed that the sufficient and necessary conditions are satisfied, if and only if the conditions in \eqref{suf_stab_baeline1} hold. 
\end{IEEEproof}



Lemma~\ref{lem_stab_baseline} and Corollary~\ref{col_stab_baseline} show the scalability and stability tradeoff imposed by the IoT network and give prior information about the network stability before conducting the iterative solution. The upper bound in Corollary~\ref{col_stab_baseline} identifies the maximum spatial intensity $\tilde{\alpha}$ device/BS/code that can be supported by cellular networks for a given traffic requirements and SINR detection threshold. Furthermore, the sufficient condition given in Lemma~\ref{lem_stab_baseline}  determines $\tilde{\alpha}$ that guarantees stable network operation. Note that $\tilde{\alpha}$ can be decreased by network densification and/or increasing the number of orthogonal channels available for IoT operation.  To study the network stability when the necessary condition is satisfied but the sufficient condition is not satisfied, the iterative solution should be utilized. Also, the network performance can only be determined via the iterative algorithm given in the next theorem.


\begin{theorem}
\label{theorem1}
The probability of being in the idle state with an empty buffer for a generic IoT device operating with the baseline transmission scheme is obtained via \bf{Algorithm 2.}

\begin{figure*}
\begin{algorithm}[H]
\small
Initialize $x_\circ$.\\
\While { $\left|x_\circ^{[i]} - x_\circ^{[i-1]} \right| \geq \epsilon $  } {
1- Calculate $p$ in  \eqref{eq:Out1} using $x_\circ^{[i]}$.\\
2- Check the stability condition. \\
\eIf{$p>a$}{
   Calculate $x_\circ^{[i+1]}$ using \eqref{eq_base}.\;
   }{
    return $x_\circ \leftarrow 0$ and calculate $p$ in  \eqref{eq:Out1} for $\bar{x}_\circ =1$. \\
    Break.\;
  }
  3- Increment $i$.} 
  \vspace{-3mm}
  return $x_\circ \leftarrow x_\circ^{[i]}$ and $p$.		
 \caption{Computation of $x_\circ$ and $p$  for the baseline scheme.}
 \normalsize
 \label{base_algo}
\end{algorithm}
\end{figure*}
\vspace{-3mm}
\begin{proof}
The proof follows from Lemma~\ref{lem:out_base} and \eqref{eq_base}.
\end{proof}
\end{theorem}

Using $x_\circ$ and $p$  from Theorem 1, the baseline IoT network can be fully characterized. For stable network, the steady state probability distribution can be obtained via~\eqref{eq:Out1}.  For stable networks, the average queue length and the average queueing delay can be obtained for Geo/Geo/1 as \cite{alfa_DTMC} 
\vspace{-1mm}
\small
 \begin{align} \label{ave_queue_baseline}
\mathbb{E}\left[Q_L\right] = \frac{a \bar{a}px_\circ}{(p-a)^2} \quad \text{and} \quad \mathbb{E}\left[W_q\right] = \frac{a \bar{a} x_\circ}{(p-a)^2}.
\end{align}
\normalsize
\vspace{-2mm}
It is worth emphasizing that \eqref{ave_queue_baseline} cannot be used directly to study the queue size and queueing delay when varying  the arrival rate. Instead, the steady state $x_\circ$ and $p$ should be first characterized via Algorithm~\ref{base_algo} for any change in the arrival rate $a$, then  \eqref{ave_queue_baseline} can applied if the network is stable.

\vspace{-3mm}
\subsection{Power-Ramping Scheme} \label{sec_ramp}
\vspace{-2mm}
In the power-ramping scheme, the transmission success probability depends on the transmission phase, and hence, a Geo/PH/1 queueing model is employed. From Fig.~\ref{fig_Markov_power_ramping}, it can be observed that each IoT device increments its power control threshold upon transmission failure, which gives the following PH type transient matrix 
\vspace{-2mm}
\small
\begin{equation} \label{S_ramp}
\bold{S}=\begin{bmatrix}
 0&\bar{p}_1& 0  & 0 &\hdots  &0\\ 
 0 & 0 &\bar{p}_2& 0 &\hdots  &0\\ 
 0& 0 & 0 & \bar{p}_3 &\hdots  &0\\
 \vdots & \vdots&  \vdots& \ddots & \ddots  &\vdots \\ 
 0 &  0&  0& 0 &\hdots  &\bar{p}_{M-1}\\
 \bar{p}_M &  0&  0& 0 &\hdots &0
\end{bmatrix}.
\end{equation}
\normalsize

Let $\mathbf{x}=[x_\circ, \bold{x}_1, \bold{x}_2,\ \hdots,\ ]$ be the stationary distribution, where $x_\circ$ represents the probability of having an idle queue and $\bold{x}_i = [x_{i,1} x_{i,2} \dots x_{i,M}]$ is the probability vector for transmission phases when of having $i$ packets in the queue. The steady state solution for the power-ramping queueing model is given in the following lemma

\begin{lemma} \label{lem_ramp1}
 Solving \eqref{eq:distribution1} with the transition matrix in \eqref{eq:trans_matrix1} and the transient matrix in \eqref{S_ramp} gives the following steady state solution  
\vspace{-2mm}
\small
\begin{align} 
x_\circ= \left( 1+\bold{C}  \left(  [\bold{I}-  a \mathbf{s}\boldsymbol{\beta} - \bar{a} \mathbf{S} - \bold{R} \bar{a} \mathbf{s}\boldsymbol{\beta} ] [\bold{I}-\bold{R}] \right)^{-1}  \mathbf{e} \right)^{-1} \quad \text{and}\quad \bold{x}_i =\left\{\begin{matrix}
 x_\circ \bold{C} [\bold{I}-  a \mathbf{s}\boldsymbol{\beta} - \bar{a} \mathbf{S} - \bold{R} \bar{a} \mathbf{s}\boldsymbol{\beta} ]   &  i=1\\ 
  {\bold{x}_1} \bold{R}^{i-1}  &  i>1
\end{matrix}\right.,
\label{eq_ramp}
\end{align}
\normalsize

\vspace{-2mm}
\noindent where $\bold{R}$ is the {rm MAM} $\bold{R}$ matrix and is given by $\bold{R}=  a \mathbf{S}[\bold{I}-a \mathbf{s}\boldsymbol{\beta} -\bar{a} \mathbf{S} -  a \mathbf{S} \mathbf{e} \boldsymbol{\beta}]^{-1}$.
\normalsize
\end{lemma}
 \begin{IEEEproof}
 $x_\circ$ and $\mathbf{x}_1$ are obtained by solving the boundary equation $\mathbf{x}_1=x_\circ\bold{C}+\mathbf{x}_1(\bold{A}_1+\bold{R}\bold{A}_2)$ and normalization condition $x_\circ+\mathbf{x}_1 [\bold{I}-\bold{R}]^{-1}\mathbf{e}=1$, where $\bold{A}_1$ and $\bold{A}_2$ are defined in \eqref{eq:trans_matrix1}. Then $\mathbf{x}_i$ in \eqref{eq_ramp} follows from the definition of the $\bold{R}$ matrix~\cite{MAM, alfa_DTMC}, which is the minimal non-negative solution of $\bold{R}=\bold{A}_0+\bold{R}\bold{A}_1+\bold{R}^2\bold{A}_2$. Since  $\bold{A}_2$ is a rank one matrix, an explicit expressions for $\bold{R}$ is obtained as  $\bold{R}=  a \mathbf{S}[\bold{I}-a \mathbf{s}\boldsymbol{\beta} -\bar{a} \mathbf{S} -  a \mathbf{S} \mathbf{e} \boldsymbol{\beta}]^{-1}$~\cite{alfa_DTMC}. 
 \end{IEEEproof}
 
Let $\Pi_m$ be the marginal probability of  using the power control threshold $\rho_m$. Following \cite{MAM, alfa_DTMC} and using the steady state solution in Lemma~\ref{lem_ramp1}, the marginal probability distribution of the phases $\boldsymbol{\Pi}=[{\Pi}_1, {\Pi}_2, \dots, {\Pi}_M]$ can be obtained as
\vspace{-3mm}
\begin{align}
\boldsymbol{\Pi}&=\mathbf{x}_1[\bold{I}-\bold{R}]^{-1}.
\label{eq:marg_ramp}
\end{align}
The marginal probability distribution $\bold{\Pi}$ is necessary for calculating the transmission success probabilities $p_m$, $\forall m$ in the power-ramping scheme, which are characterized via stochastic geometry in the following lemma.

\begin{lemma} \label{lem:out_ramp}
In the depicted IoT network with power-ramping scheme, the transmission success probability for an IoT device operating with the power control threshold $\rho_m $ is given by
\vspace{-1mm}
\small
    \begin{align}\label{eq:Out_ramp}
 p_m  &  \;\;\; \approx    \;\; \exp{\Bigg\{ - \frac{\sigma^2 \theta}{\rho_m} \Bigg\}} \mathlarger{\prod}\limits_{k=1}^{M}   \frac{ \exp\Bigg\{ -  \frac{2 \Pi_k \tilde{\alpha} {\theta \rho_{k,m}} }{(\eta-2)} \; \; _2 F_1\left(1,1-2/\eta,2-2/\eta,-{\theta \rho_{k,m}}  \right) \Bigg\}}{\left({ 1+ \frac{{\theta \rho_{k,m}}}{\left(1+{\theta \rho_{k,m}} \right)}\frac{\Pi_k \tilde{\alpha}}{c}}\right)^{c} } \notag \\
 & \overset{(\eta=4)}{=}  \exp{\Bigg\{ - \frac{\sigma^2 \theta}{\rho_m} \Bigg\}} \mathlarger{\prod}\limits_{k=1}^{M}     \frac{ \exp\Bigg\{  -  { \Pi_k \tilde{\alpha} \sqrt{\theta \rho_{k,m}} } \; \; \arctan\left(\sqrt{\theta \rho_{k,m}}  \right) \Bigg\}}{\left({ 1+ \frac{{\theta \rho_{k,m}}}{\left(1+{\theta \rho_{k,m}} \right)}\frac{\Pi_k \tilde{\alpha}}{c}}\right)^{c} },
    \end{align}
\normalsize
  where $\rho_{m,k} = \frac{\rho_m}{\rho_k}$, $c=3.575$, and the approximation is due to the employed approximate PDF of the $\mathbb{R}^2$ PPP Voronoi cell area in \eqref{pdf_users} and Approximation 1.
\end{lemma}
\vspace{-4mm}
\begin{IEEEproof}
See Appendix \ref{sec:AppB}.
\end{IEEEproof}

The interdependences between $\bold{\Pi}$ in \eqref{eq:marg_ramp}, $\bold{S}$ in \eqref{S_ramp}, and $p_m$ in \eqref{eq:Out_ramp} clearly show the causality problem in the power-ramping scheme as  discussed in Section~\ref{causal}. Similar to the baseline scheme, we can study the necessary and sufficient conditions for network stability in the power-ramping scheme prior  to applying the iterative solution to find $p_m$ and $\bold{\Pi}$. Let {$\bold{A}=\bold{A}_0+ \bold{A}_1 +\bold{A}_2 = \mathbf{s}\boldsymbol{\beta}+\mathbf{S}$} and  $\boldsymbol{\pi}=[\pi_1, \pi_2, \dots, \pi_M]$, then solving the system 
\vspace{-1mm}
\small
\begin{equation} \label{sabb_pramp}
\boldsymbol{\pi}  \bold{A}=\boldsymbol{\pi}  \quad \text{and} \quad  \boldsymbol{\pi}  \mathbf{e}=1,
\end{equation}
\normalsize 
leads to 
\small
\begin{equation} \label{sabb_pramp1}
\pi_{m+1}= \pi_{m} \bar{p}_m, \quad \text{where} \quad \pi_1 = \left(1+\sum\limits_{i=1}^{M} \prod_{m=1}^i \bar{p}_m \right)^{-1}
\end{equation} 
\normalsize
\noindent where $p_m$ is given in \eqref{eq:Out_ramp}.  The power-ramping queueing system is stable if and only if
\vspace{-1mm}
\small
\begin{align}
\bar{a} \boldsymbol{\pi}  \mathbf{s}\boldsymbol{\beta}  \mathbf{e}> a \boldsymbol{\pi} \mathbf{S} \mathbf{e}.
\label{stability_power_ramping}
\end{align}
\normalsize
The necessary condition is obtained via \eqref{stability_power_ramping} with $p_m$ in \eqref{eq:Out_ramp}  at $x_\circ=1-a$ and $\Pi_1=a$, which represent the mildest traffic and interference. Furthermore, the sufficient condition is obtained via \eqref{stability_power_ramping} with $p_m$ in \eqref{eq:Out_ramp} at  $\Pi_M =1$, which represent the highest traffic and strongest interference. Similar  to the baseline scheme, the necessary and sufficient conditions can be exploited to find bounds on the intensity of IoT deceives that a network can support.  
The steady state solution for the power-ramping scheme is obtained via the iterative algorithm given in the following theorem.

\begin{theorem}
\label{theorem2}
The marginal steady state probability vector $\bold{\Pi}$ for a generic IoT device operating with the power-ramping transmission scheme is obtained via \bf{Algorithm 3.}
 
\begin{figure*}
\begin{algorithm}[H]
\small
Initialize $x_\circ$ and $\bold{\Pi}$ such that $x_\circ + \bold{\Pi} \times \bold{e} = 1$.\\
\While { $\left|\bold{\Pi}^{[i]} - \bold{\Pi}^{[i-1]} \right| \geq \epsilon $  } {
1- Calculate $p_m$, $\forall m$ in  \eqref{eq:Out_ramp} using $\bold{\Pi}^{[i]}$.\\
2- Construct $\bold{S}$ using $p_m$ as in \eqref{S_ramp}. \\
3- Construct  $\boldsymbol{\pi}=[\pi_1, \pi_2, \dots, \pi_M]$, where $\pi_{m+1}= \pi_{m} \bar{p}_m$ and  $\pi_1 = \left(1+\sum\limits_{i=1}^{M} \prod_{m=1}^i \bar{p}_m \right)^{-1}$.\\
4- Check the stability condition. \\
\eIf{$\boldsymbol{\pi} \bar{a} \mathbf{s}\boldsymbol{\beta}  \mathbf{e}>\boldsymbol{\pi} a \mathbf{S} \mathbf{e}$}{
   Calculate $\bold{x}_1$ from Lemma~\ref{lem_ramp1} and $\bold{\Pi}^{[i+1]}$ from \eqref{eq:marg_ramp}.\;
   }{
    Set $x_\circ \leftarrow 0$. \\
    Solve the system $\bold{\Pi} \mathbf{P} = \bold{\Pi}$ and $\bold{\Pi} \bold{e} = 1$ with the transition matrix in \eqref{eq:distribution2}.}
  5- Increment $i$.}
  \vspace{-3mm}
  return $\bold{\Pi} \leftarrow \bold{\Pi}^{[i]}$, $x_\circ \leftarrow 1-\bold{\Pi}\bold{e} $, and $p_m$.		
 \caption{Computation of $x_\circ$ and $p$  for the power-ramping scheme.}
 \label{ramp_algo}
 \normalsize
\end{algorithm}
\end{figure*}
\vspace{-2mm}
\begin{proof}
The proof follows from Lemma~\ref{lem_ramp1} and Lemma~\ref{lem:out_ramp}.
\end{proof}
\end{theorem}
\vspace{-3mm}
Using $x_\circ$, $\bold{\Pi}$ and $p_m$  from Theorem 2, the IoT network operating with the power-ramping scheme can be fully characterized.  Applying the law of total probability, the average transmission success probability conditioned on  that the IoT device is active can be given by

\small
\begin{align}\label{outage_ramping}
 P_{success}= \mathbb{E}[{p}_m] =\displaystyle{\sum\limits_{m=1}^{M} \frac{\boldsymbol{{\Pi}}_m}{1-x_\circ} p_m}.
\end{align} 
\normalsize
From \eqref{ave_retrans}, the mean number of retransmissions for each successful packet delivery is given by $\mathcal{D}= (1-x_\circ)/\displaystyle{\sum\limits_{m=1}^{M} \boldsymbol{{\Pi}}_m p_m}$. 
For stable network, the steady state probability distribution and the $\bold{R}$ matrix can be obtained via~Lemma~\ref{lem_ramp1}. Also, by using \eqref{ave_queue}, the average queue length for stable network is given by
\vspace{-1mm}
\small
\begin{align} \label{ave_queue_power}
\mathbb{E}\left[Q_L\right] &= (\mathbf{x}_1 +2\mathbf{x}_2 + 3\mathbf{x}_3 + \cdots)\bold{e} = \mathbf{x}_1(1 +2\bold{R} + 3\bold{R} + \cdots)\bold{e}\notag \\
 &=\bold{x}_1 (\bold{I}-\bold{R})^{-2}\bold{e}.
\end{align}
\normalsize
where \eqref{ave_queue_power} follows from the fact that $\mathbf{R}$ has a spectral radius less than one  \cite{alfa_DTMC}. The average queuing delay is given by solving \eqref{wait_gen} with $\bold{S}$ given in \eqref{S_ramp} along with $\bold{x}$ and $p_m$ obtained through Algorithm~\ref{ramp_algo}.

\vspace{-2mm}
\subsection{Backoff Scheme}

In the backoff scheme, the IoT devices defer their transmissions and go to backoff upon transmission failures. From Fig.~\ref{fig_Markov_power_ramping}, it can be observed that there are deterministic and probabilistic backoff states in which the PH type transient matrix can be represented as

\vspace{-2mm}

\begin{equation}
\small
\mathbf{S} = \begin{bmatrix}
0 & \overline{p} & 0 & \cdots  & 0 & 0\\ 
0 & 0 & 1 &  \cdots  & 0& 0\\ 
\vdots  & \vdots   & \vdots  & \ddots   & \vdots & \vdots \\ 
q & 0 & 0 &  0 & 0& \overline{q} \\ 
q & 0 & 0 & 0 &  0& \overline{q} \\ 
\end{bmatrix}.
\normalsize
\label{S_back}
\end{equation}

Let  $\mathbf{x}=[x_\circ, x_{1,1}, x_{1,2}, \dots, x_{i,j-1}, x_{i,j}, x_{i,j+1} \dots]$ be the stationary distribution, where $x_\circ$ represents the probability of having an idle queue, $x_{i,1}$ represents the probability of being in the transmission phase, and $x_{i,j}$ for $i\geq1$ and $j>1$ represents the probability of being in one of the backoff phases (i.e., having non-empty buffer but not transmitting). The steady state solution for the backoff scheme can be obtained via Lemma~\ref{lem_ramp1} but with $\bold{S}$ given in \eqref{S_back}.

Using the same definition for $\bold{\Pi}$ as in Section~\ref{sec_ramp}, it is noticed that $\Pi_1$ is the probability of being in transmission state and $\Pi_n$ $n>1$ is the probability of being in the backoff phase. The marginal distribution $\bold{\Pi}$ for the backoff scheme can be obtained via \eqref{eq:marg_ramp} but with $\bold{S}$ given in \eqref{S_back}. As discussed earlier, the marginal distribution $\bold{\Pi}$ is required to obtain the transmission success probability, which is given in the following lemma.

\begin{lemma} \label{lem:out_back}
The transmission success probability in the depicted IoT network with backoff transmission scheme is given by 
\small
\begin{align}\label{eq:Out_back}
  {p}   &  {\approx} \frac {\exp\left\{- \frac{\sigma^2 \theta}{\rho} -  \frac{2 \theta \Pi_1 \tilde{\alpha}}{(\eta-2) } \;{}_2F_1\left(1,1-2/\eta,2-2/\eta,-\theta\right) \right\}}{ \left(  1+  \frac{\theta \Pi_1 \tilde{\alpha}}{(1+\theta) c}\right)^{c} }
  \overset{(\eta=4)}{=} \frac{\exp\left\{- \frac{\sigma^2 \theta}{\rho} - \Pi_1 \tilde{\alpha} \sqrt{\theta}\arctan\left({\sqrt{\theta}}\right)   \right\}}{ \left(  1+  \frac{\theta \Pi_1 \tilde{\alpha}}{(1+\theta)c}\right)^{c} }.
\end{align}
\normalsize
   where $c=3.575$ and the approximation is due to the employed approximate PDF of the $\mathbb{R}^2$ PPP Voronoi cell area in \eqref{pdf_users} and Approximation 1.
\end{lemma}
\vspace{-1mm}
\begin{IEEEproof}
See Appendix \ref{sec:AppA}.
\end{IEEEproof}

%
%

The condition for network stability for the backoff scheme can be determined following the same methodology that is used in Section~\ref{sec_ramp}. Specifically, let $\boldsymbol{\pi} = [\pi_1, \pi_2, \dots, \pi_{N+1}, \pi_q]$, then solving \eqref{sabb_pramp} with $\bold{S}$ given in \eqref{S_back} leads to  $\pi_i=\bar{p} \pi_1$ for $i\in\{2,\dots,N+1\}$ and $\pi_q=\pi_1\bar{p} \left(\frac{1}{q}-1\right)$, where $\pi_1 = \left(1+(N-1)\bar{p}+\frac{\bar{p}}{q}\right)^{-1}$ and $p$ is given in \eqref{eq:Out_back}.  Consequently, the necessary and sufficient conditions for the backoff scheme can be respectively obtained via $(\bar{a} \boldsymbol{\pi}  \mathbf{s}\boldsymbol{\beta}  \mathbf{e}> a \boldsymbol{\pi} \mathbf{S} \mathbf{e})$ by setting $\Pi_1=a$ and $\Pi_1=1$ in $p$. The steady state probabilities for the backoff scheme are obtained via the iterative algorithm given in the following theorem.

\begin{theorem}
\label{theorem3}
The marginal steady state probability vector $\bold{\Pi}$ for a generic IoT device operating with the backoff transmission scheme is obtained via \bf{Algorithm \ref{back_algo}}.
 
\begin{algorithm}[H]
\small
Initialize $x_\circ$ and $\bold{\Pi}$ such that $x_\circ + \bold{\Pi} \times \bold{e} = 1$.\\
\While { $\left|\bold{\Pi}^{[i]} - \bold{\Pi}^{[i-1]} \right| \geq \epsilon $  } {
1- Calculate $p$ in  \eqref{eq:Out_back} using ${\Pi}_1^{[i]}$.\\
2- Construct $\bold{S}$ using $p$ as in \eqref{S_back}. \\
3- Construct $\boldsymbol{\pi} = [\pi_1, \pi_2, \dots, \pi_{N+1}, \pi_q]$ such that  $\pi_i=\bar{p} \pi_1$ for $i\in\{2,\dots,N+1\}$ and $\pi_q=\pi_1\bar{p} \left(\frac{1}{q}-1\right)$, where $\pi_1 = \left(1+(N-1)\bar{p}+\frac{\bar{p}}{q}\right)^{-1}$.\\
4-Check the stability condition. \\
\eIf{$\bar{a}  \boldsymbol{\pi} \mathbf{s}\boldsymbol{\beta}  \mathbf{e}>a \boldsymbol{\pi}  \mathbf{S} \mathbf{e}$}{
   Calculate $\bold{x}_1$ from Lemma~\ref{lem_ramp1}  and $\bold{\Pi}^{[i+1]}$ from \eqref{eq:marg_ramp}.\;
   }{
    Set $x_\circ \leftarrow 0$ and solve the system $\bold{\Pi} \mathbf{P} = \bold{\Pi}$ and $\bold{\Pi} \bold{e} = 1$ with the transition matrix in \eqref{eq:distribution2}.
  }
  5- Increment $i$.}\;  	
    \vspace{-3mm}
return $\bold{\Pi} \leftarrow \bold{\Pi}^{[i]}$, $x_\circ \leftarrow 1-\bold{\Pi}\bold{e} $, and $p$.		
 \caption{Computation of $x_\circ$ and $p$  for the backoff scheme.}
 \label{back_algo}
\end{algorithm}
\normalsize
\vspace{-5mm}
\begin{proof}
Similar to Theorem 2.
\end{proof}
\end{theorem}

\vspace{-1mm}
Using $x_\circ$, $\bold{\Pi}$ and $p$  from Theorem 3, the IoT network operating with the backoff scheme can be fully characterized. The probability of successful packet transmission is given by $P_success=\mathbb{E}[{p}_m]=\frac{\Pi_1 p}{1-x_\circ}$, where the expectation accounts just for the IoT devices that transmit i.e. not in backoff state(s). From \eqref{ave_retrans}, the mean number of retransmissions for each successful packet delivery is given by:
\vspace{-8mm}

\begin{align}\label{delay_back}
\small
 \mathcal{D}= \frac{1-x_\circ}{\Pi_1 p}.
\normalsize
\end{align} 
For stable network, the steady state probability distribution and the $\bold{R}$ matrix can be obtained via~Lemma~\ref{lem_ramp1}. Also, the average queue length and the waiting time for the stable network operating with the backoff scheme can be obtained via \eqref{ave_queue_power} and \eqref{wait_gen}, respectively, with $\bold{S}$ given in \eqref{S_back} along with $\bold{x}$ and $p$ obtained through Algorithm~\ref{back_algo}.

As shown in Fig.~\ref{fig_Markov_backoff}, the size of deterministic backoff slots is determined by $N$ and the probabilistic backoff is parametrized by $q$. Consequently, $N$ and $q$ are two fundamental design parameters in the backoff scheme. The transmission probability $\Pi_1$ can be controlled by manipulating  $N$ and $q$, which impose a tradeoff between the transmission success probability and the probability of being in transmission. Particularly, selecting a large $N$ or small $q$ lead to a conservative spectrum access with high transmission success probability (i.e., high $p$), and vice versa. For the optimal selection of $N$ and $q$, we formulate the following problem 
\begin{equation}
\small
\begin{aligned}
& \underset{N,q}{\text{minimize}}
& & \mathbb{E}[W_q]  \\
& \text{subject to}
& & N \in \mathbb{Z} & & , & & 0 \leq q \leq 1 
\end{aligned}
\normalsize
\label{optimization}
\end{equation}
in which the objective is to minimize the queue waiting time, which is given in \eqref{wait_gen}. Due to space constraints, we do not delve into the analysis of  \eqref{optimization} and employ a straightforward exhaustive search solution. It is worth mentioning that the feasible set for the employed exhaustive search contains all paris  $(N,q)$ that lead to a stable queue performance. In the case where none of the combinations of $N$ and $q$ leads to a stable network, the optimal values of $N$ and $q$ are obtained by replacing $\mathbb{E}[W_q]$ with $\mathcal{D}$ in \eqref{optimization} such that the objective is changed to minimizing the average number of retransmissions given in \eqref{delay_back}.

\vspace{-2mm}
\section{Numerical Results \& Simulations}\label{sec:Results}
At first, we compare the proposed analysis with independent system level simulations. It is important to note that the simulation is used to verify the stochastic geometry analysis for the transmission success probabilities, which incorporate Approximation 1 related to the spatial correlation between the transmission powers of devices as well as the approximation of PDF of the Voronoi cell area while calculating the distribution of the number of users in the cell. On the other hand, the queueing analysis is exact, and hence, is embedded into the simulation. In each simulation run, the BSs and IoT devices are realized over a 100 km$^2$ area via independent PPPs. Each IoT device is associated to its nearest BS and employs channel inversion power control. The collected statistics are taken for devices located within 1 km from the origin to avoid the edge effects.  Unless otherwise stated, we choose $a=0.1$, $x_\circ^{[0]}=.75$, { $\tilde{\alpha} =1,\; 4,\; \text{and}\; 8$} device/BS/code,\footnote{For 10 BS/km$^{2}$ and 64 code per BS, the chosen values of $\tilde{\alpha}$ correspond to $\mathcal{U} =640,\; 2560,\; \text{and}\; 5120$ device/km$^2$.} $\eta =4$, $\rho =-90$ dBm, $\sigma^2 =-90$ dBm, and $ -20 \leq \theta \leq 0$ dB.    For the power-ramping scheme, the values of  $\rho_m$ are chosen to vary from $-90$ dBm to $-70$ dBm with $4$ dBm resolution (i.e with a maximum number of retransmissions of 5). For the backoff scheme, the values of $N$ and $q$ are obtained via \eqref{optimization} for every value of $\theta$.

 \begin{figure*}

	\begin{center}
	  \subfigure[Baseline Scheme.]{\label{fig:out1}\includegraphics[width=2.1in]{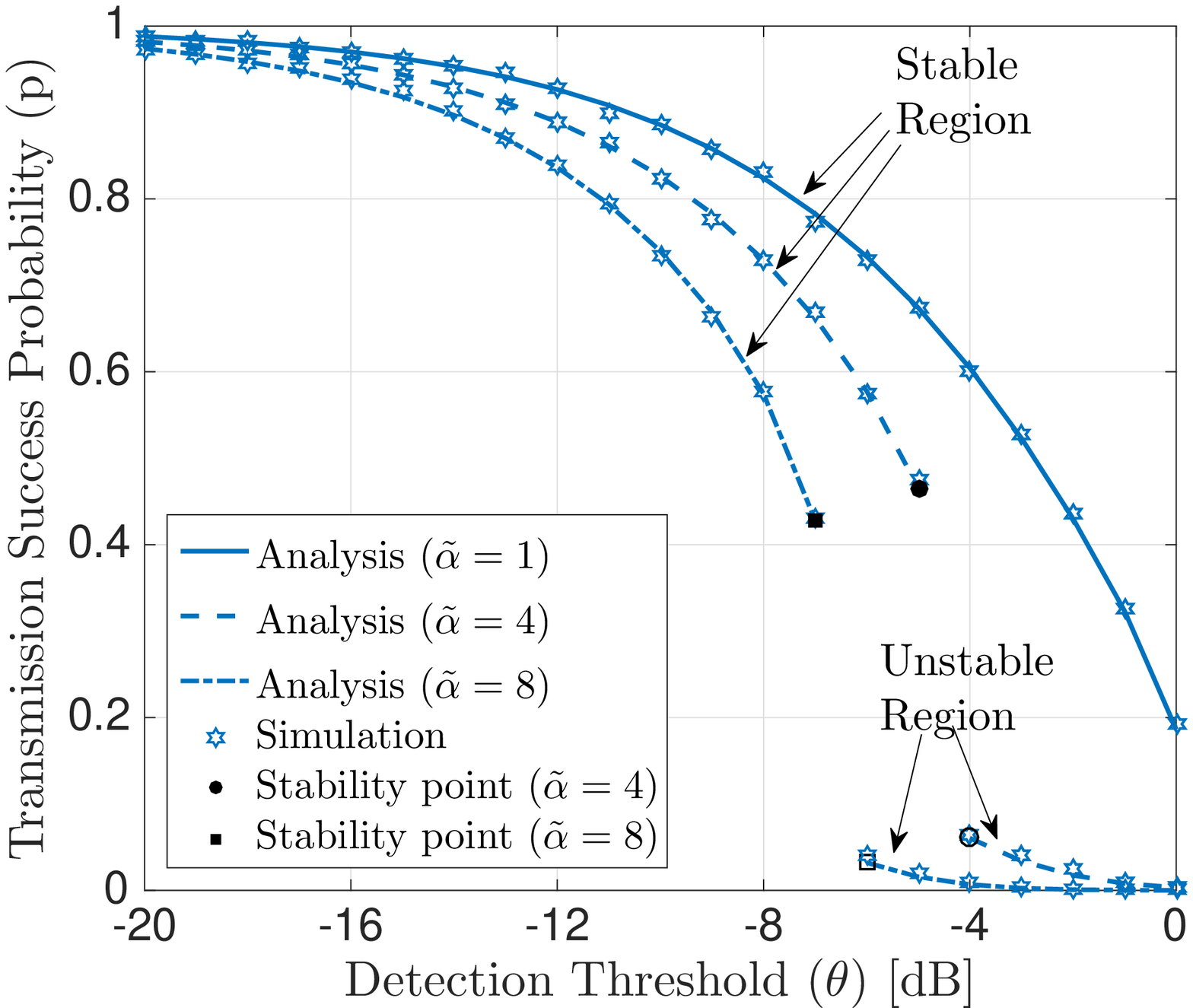}}
	  \subfigure[Power-ramping Scheme.]{\label{fig:out2}\includegraphics[width=2.1in]{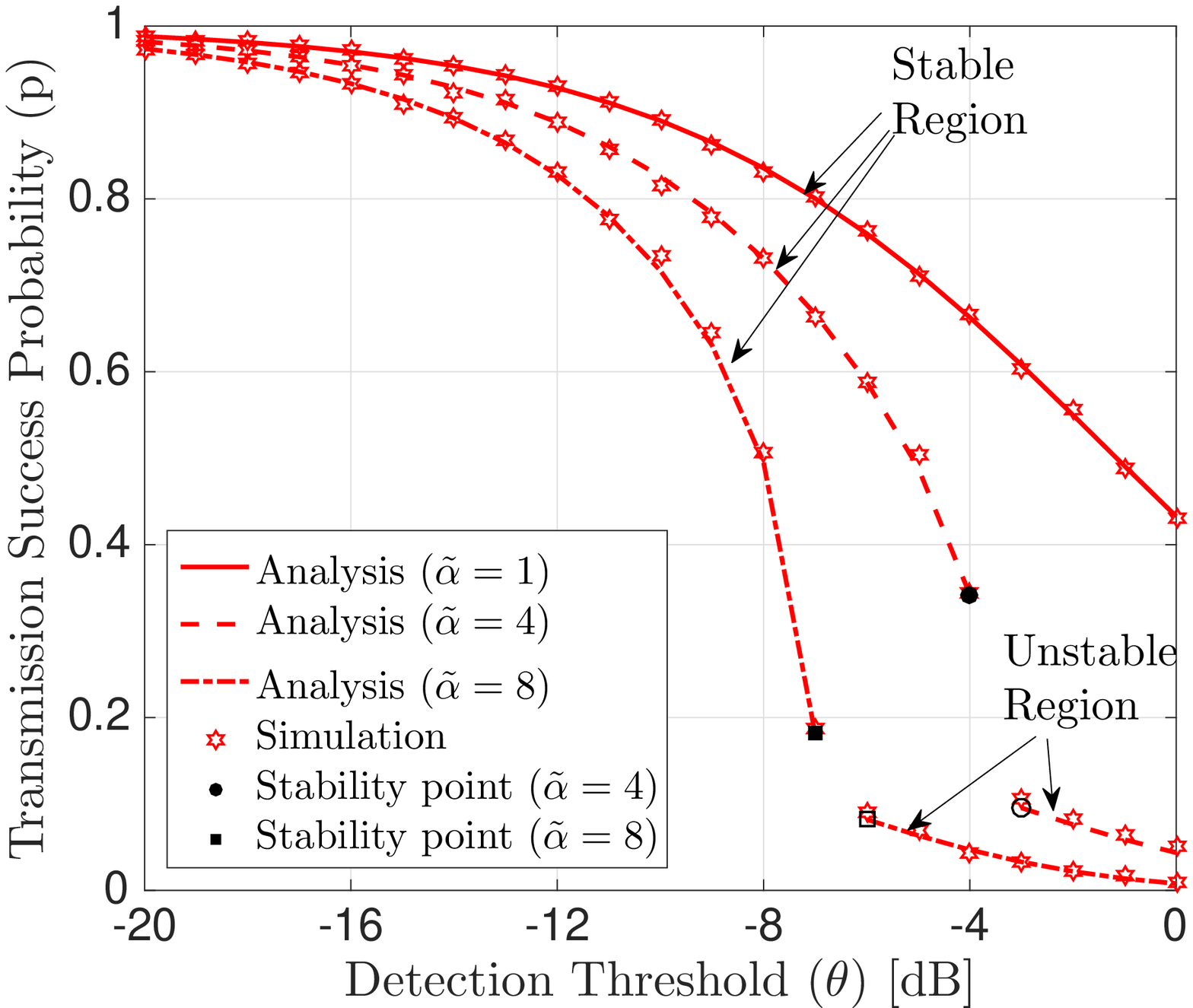}}
		\subfigure[Backoff Scheme.]{\label{fig:out3}\includegraphics[width=2.1in]{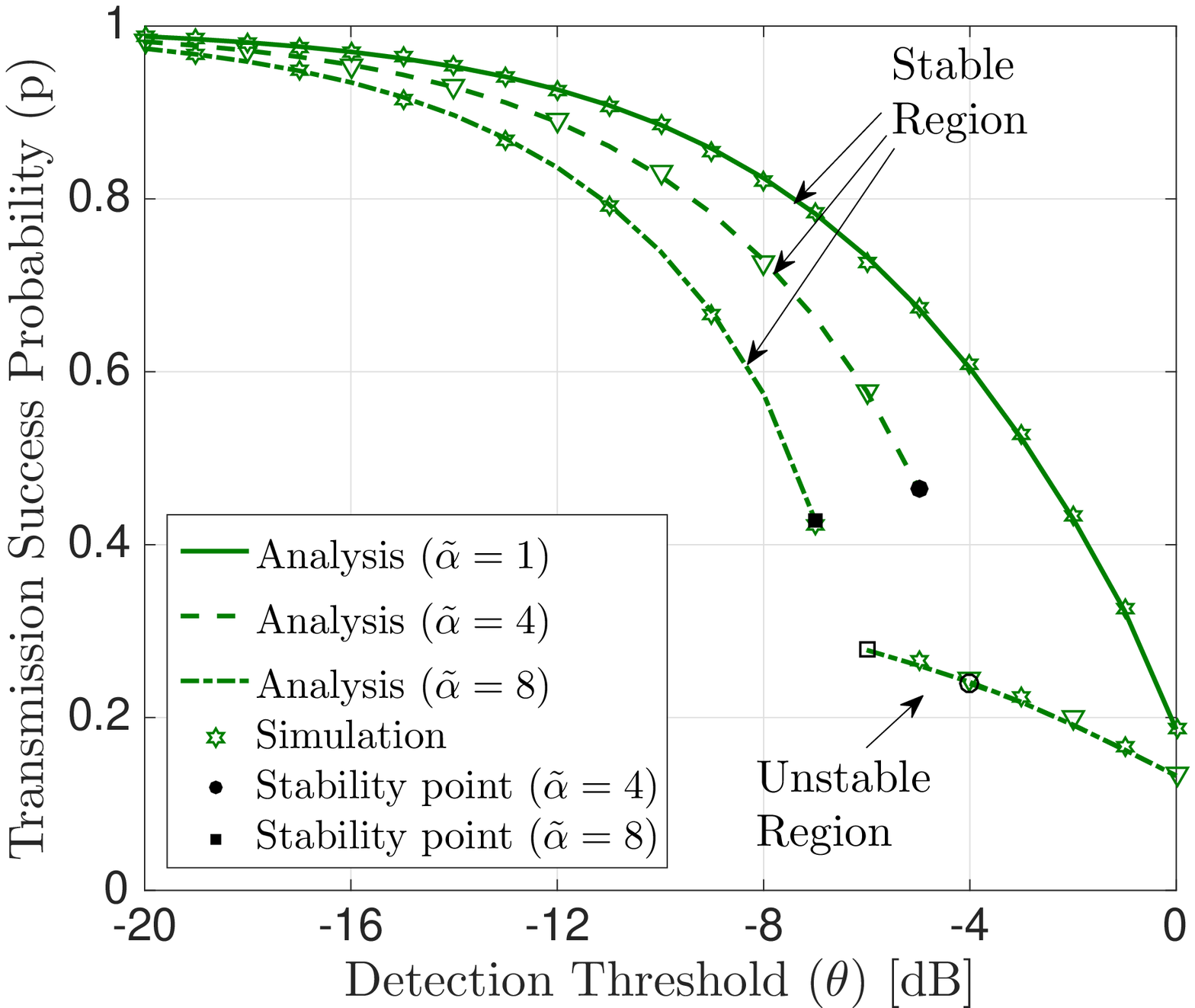}}
	\end{center}
	\vspace{-5mm}
	\caption{ Transmission success probability for the three schemes}
	\label{fig:outage}

\vspace{-2mm}
	\begin{center}
	  \subfigure[Baseline Scheme.]{\label{fig:out1}\includegraphics[width=2.1in]{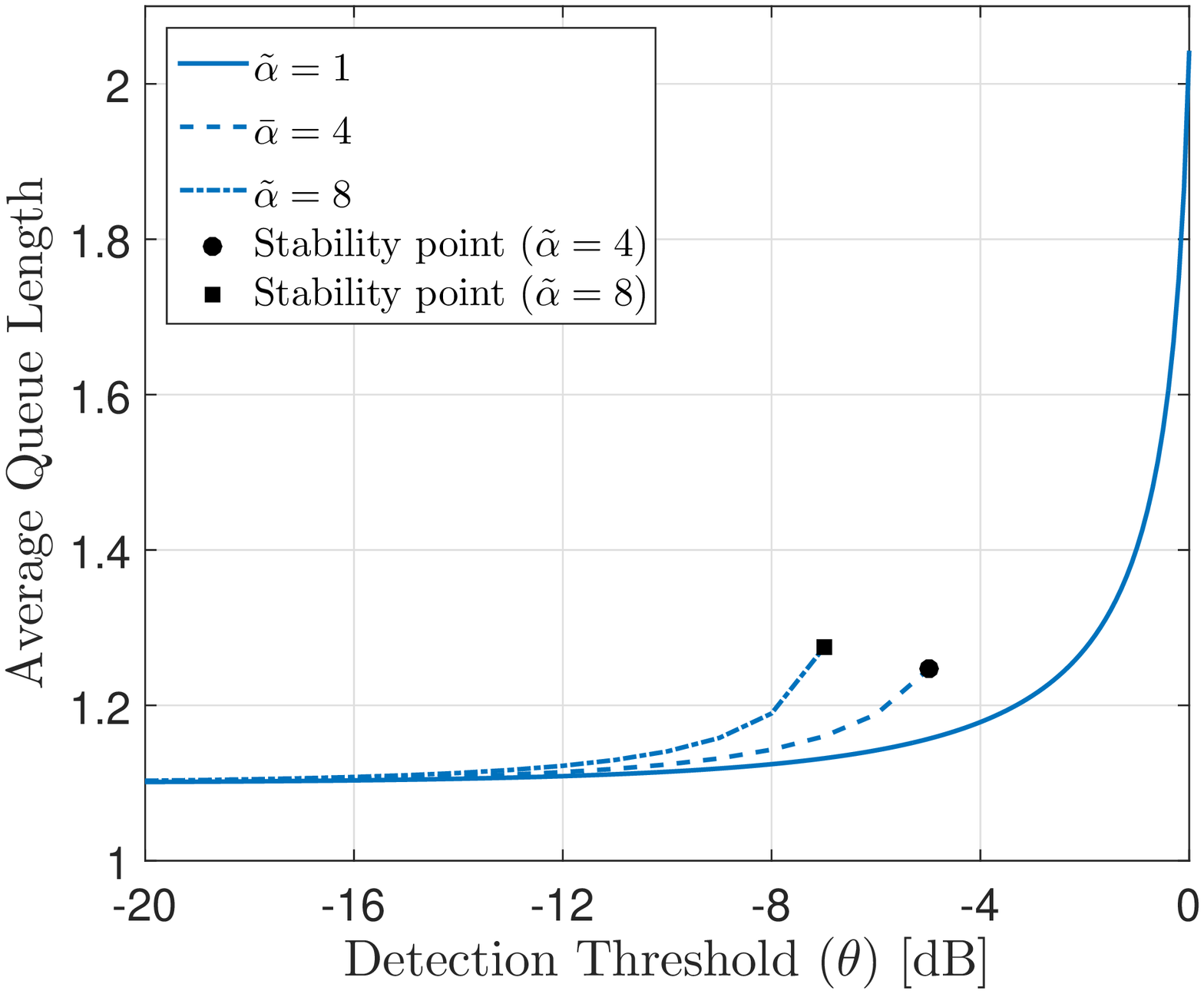}}
	  \subfigure[Power-ramping Scheme.]{\label{fig:out2}\includegraphics[width=2.1in]{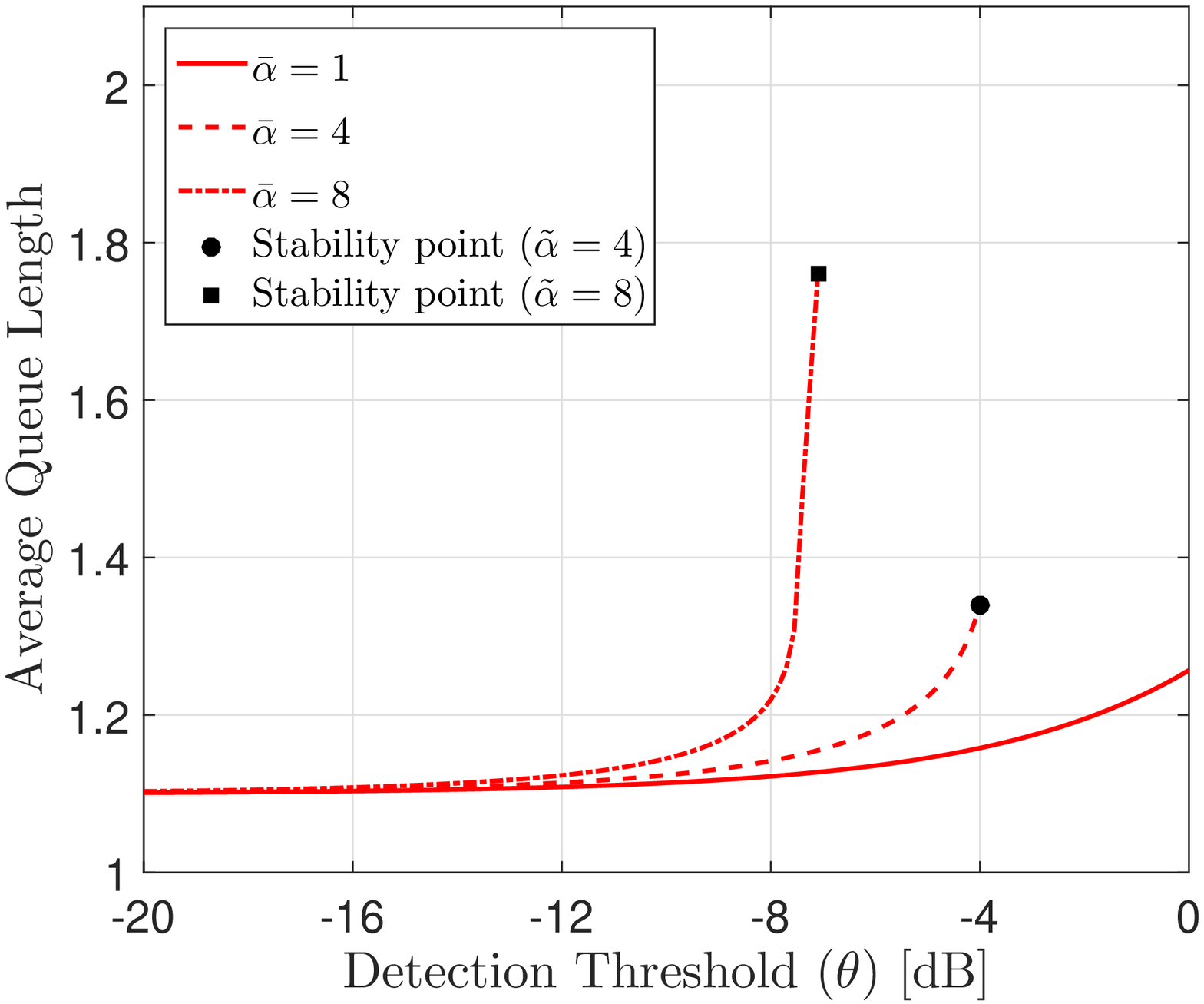}}
		\subfigure[Backoff Scheme.]{\label{fig:out3}\includegraphics[width=2.1in]{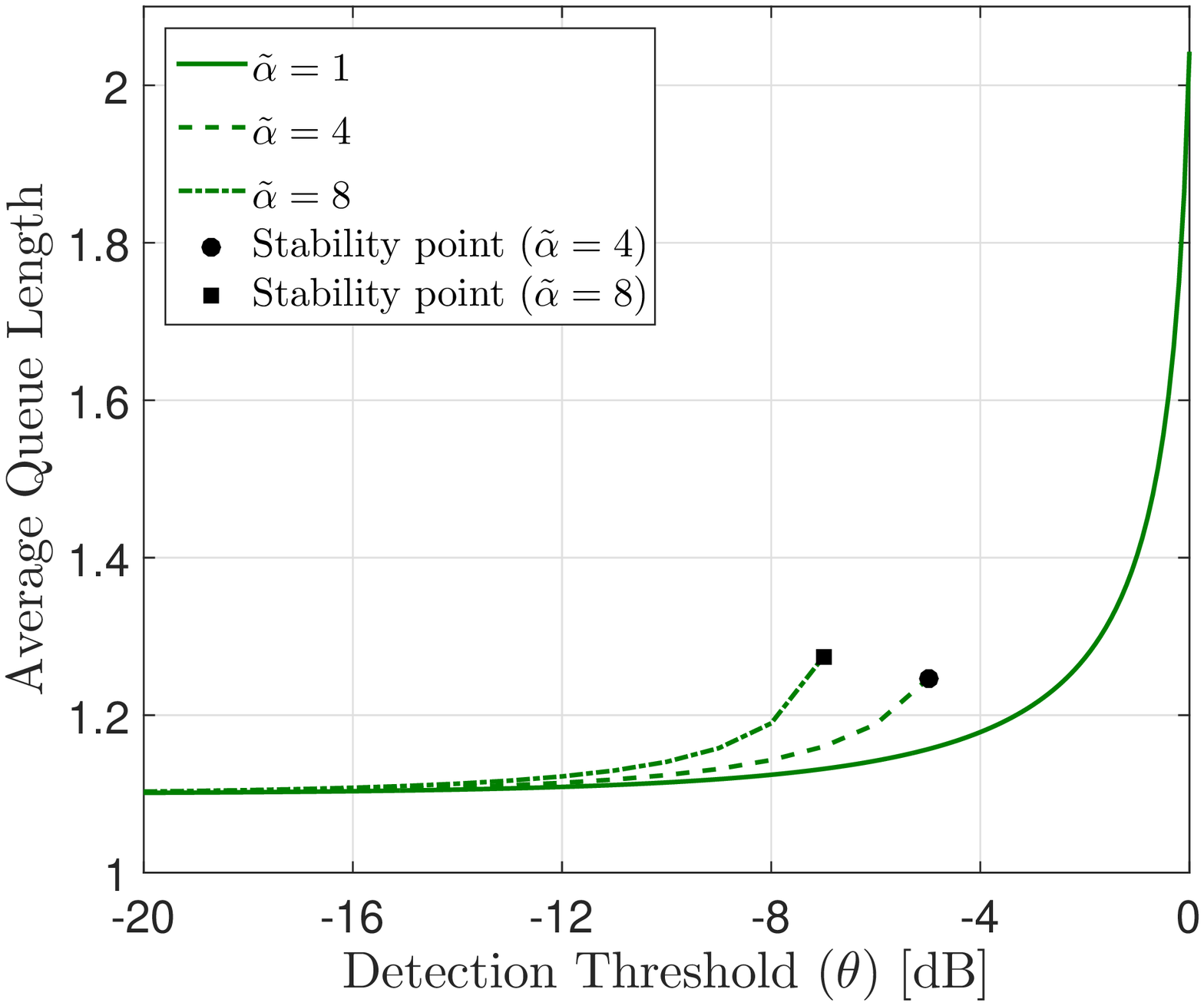}}
	\end{center}
	\vspace{-5mm}
	\caption{ Average queue length in the device's buffer for the three schemes.}
	\label{fig:queue_length}
\vspace{-5mm}
 \hrulefill	
\end{figure*}

\begin{figure*}
\vspace{-2mm}
	\begin{center}
	  \subfigure[Baseline Scheme.]{\label{fig:out1}\includegraphics[width=2.1in]{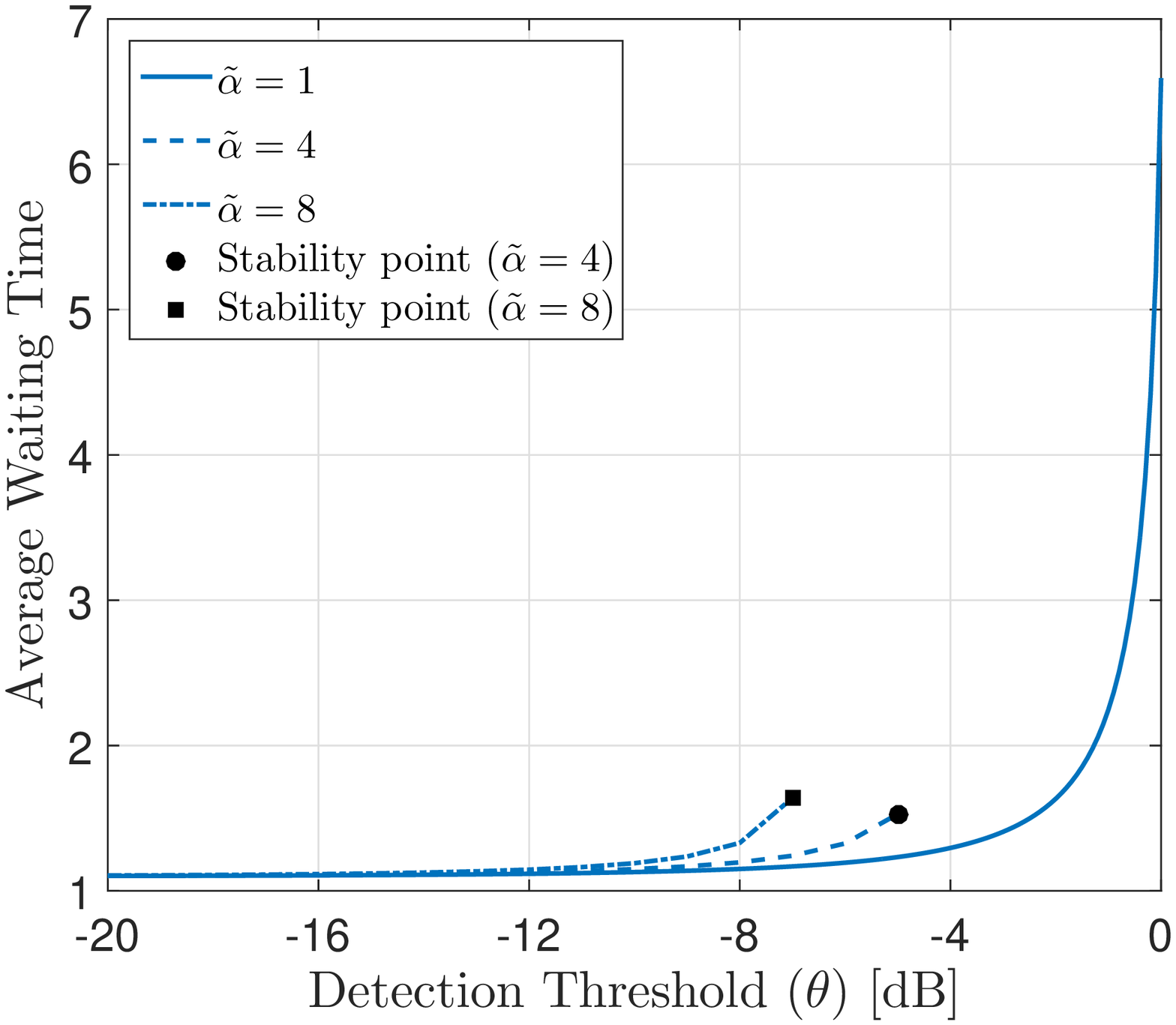}}
	  \subfigure[Power-ramping Scheme.]{\label{fig:out2}\includegraphics[width=2.1in]{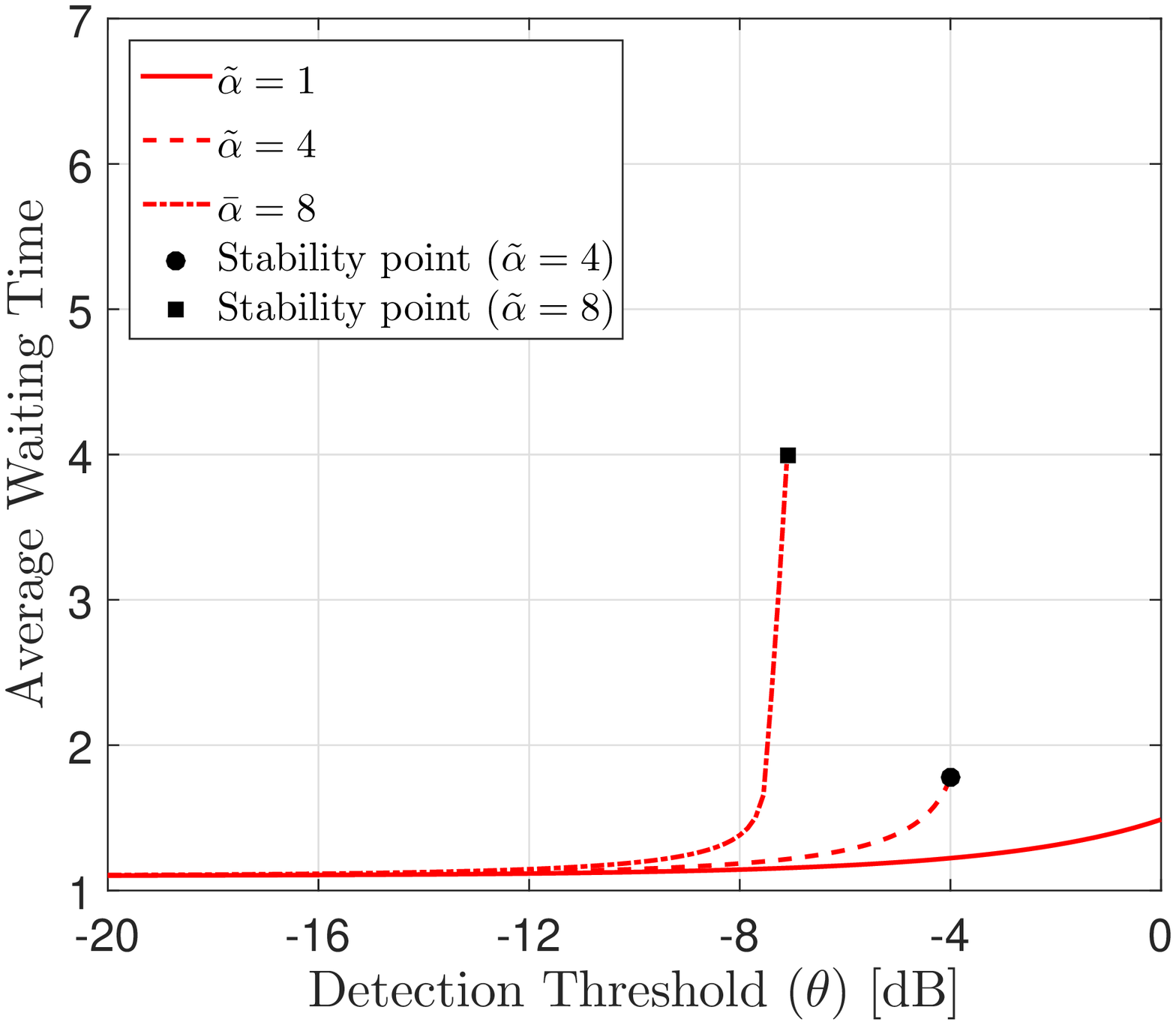}}
		\subfigure[Backoff Scheme.]{\label{fig:out3}\includegraphics[width=2.1in]{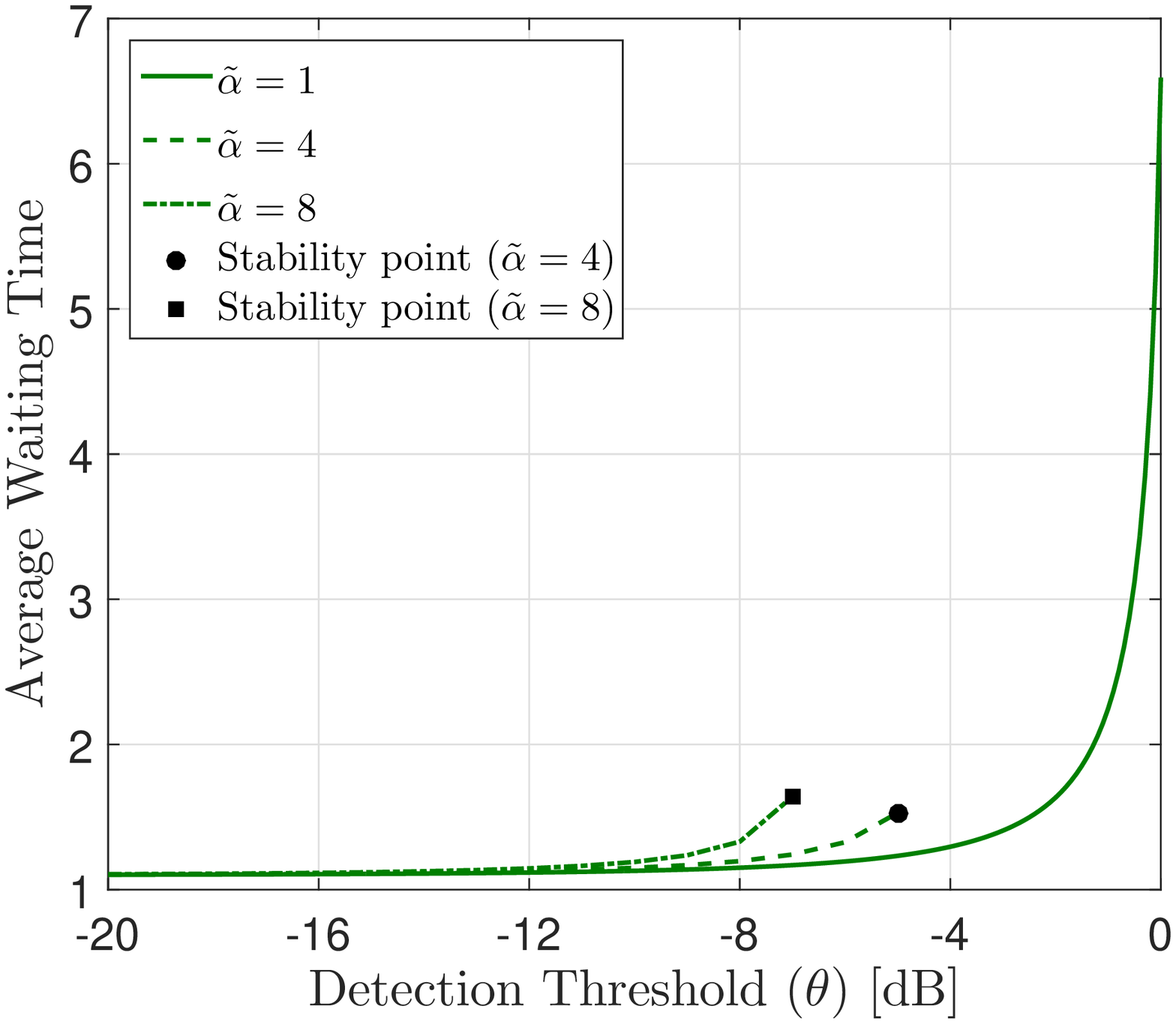}}
	\end{center}
	\vspace{-5mm}
	\caption{ Average waiting time in the queue for the three schemes.}
	\label{fig:waiting_time}

\vspace{-2mm}
	\begin{center}
	  \subfigure[Baseline Scheme.]{\label{fig:out1}\includegraphics[width=2.1in]{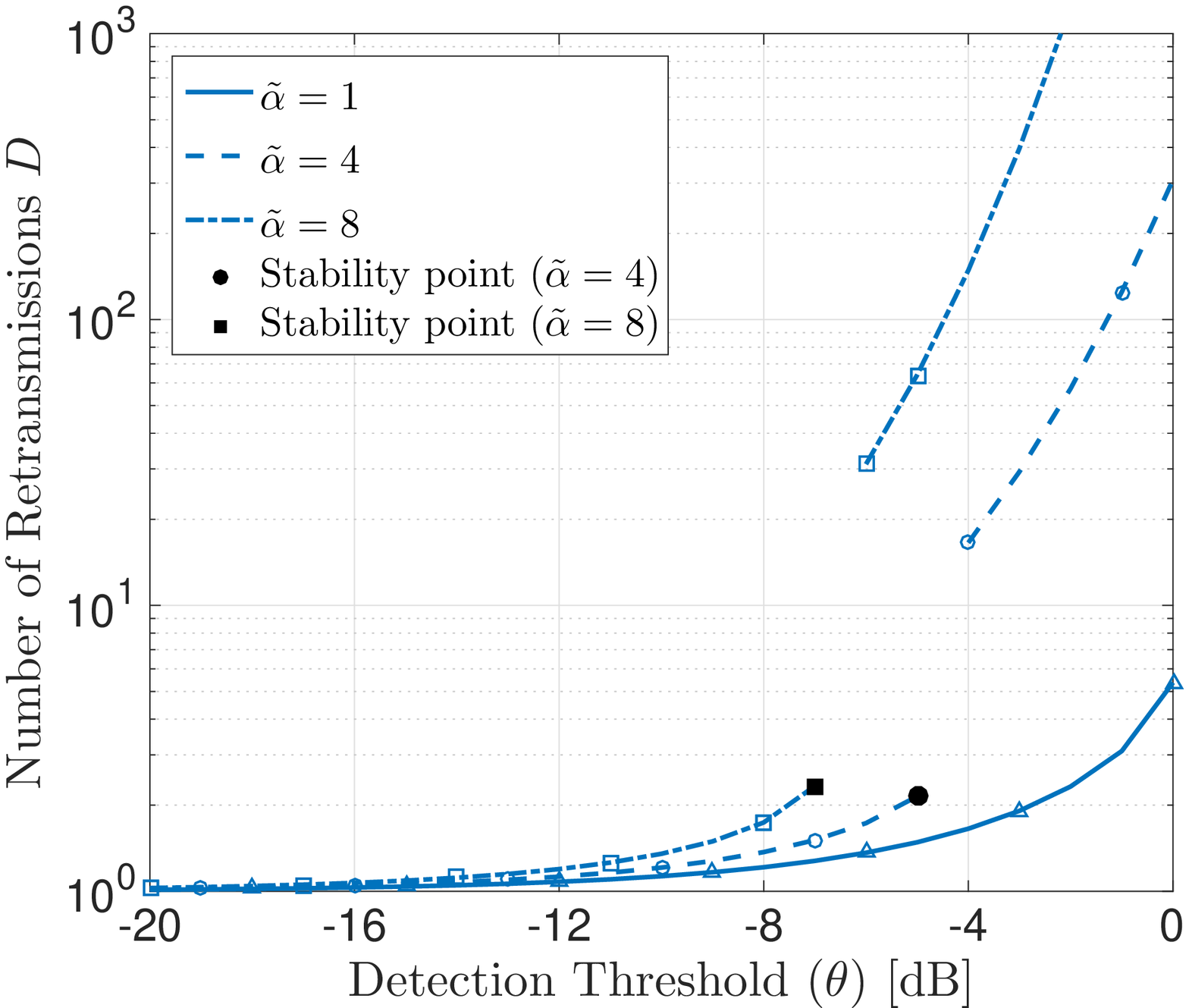}}
	  \subfigure[Power-ramping Scheme.]{\label{fig:out2}\includegraphics[width=2.1in]{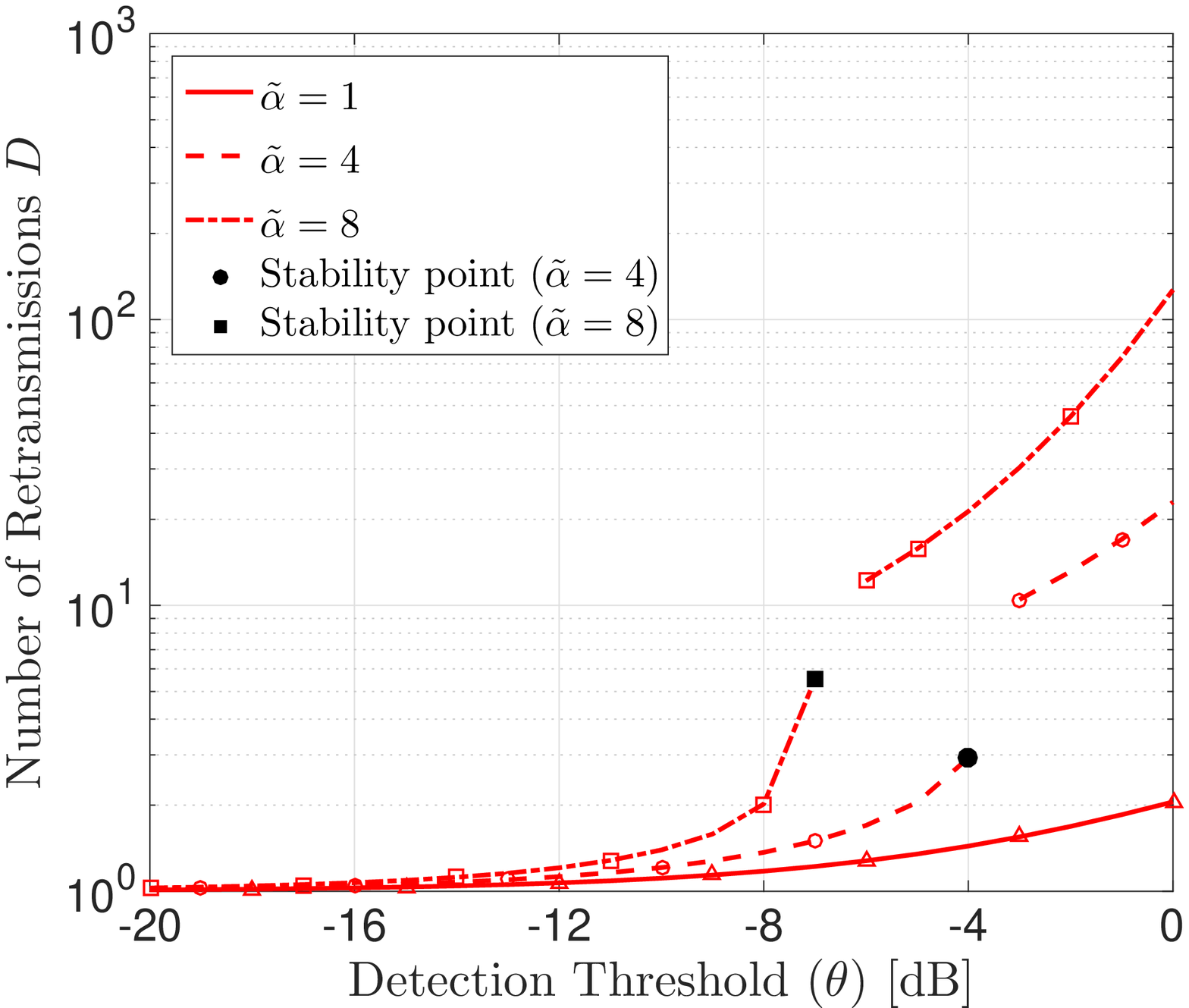}}
		\subfigure[Backoff Scheme.]{\label{fig:out3}\includegraphics[width=2.1in]{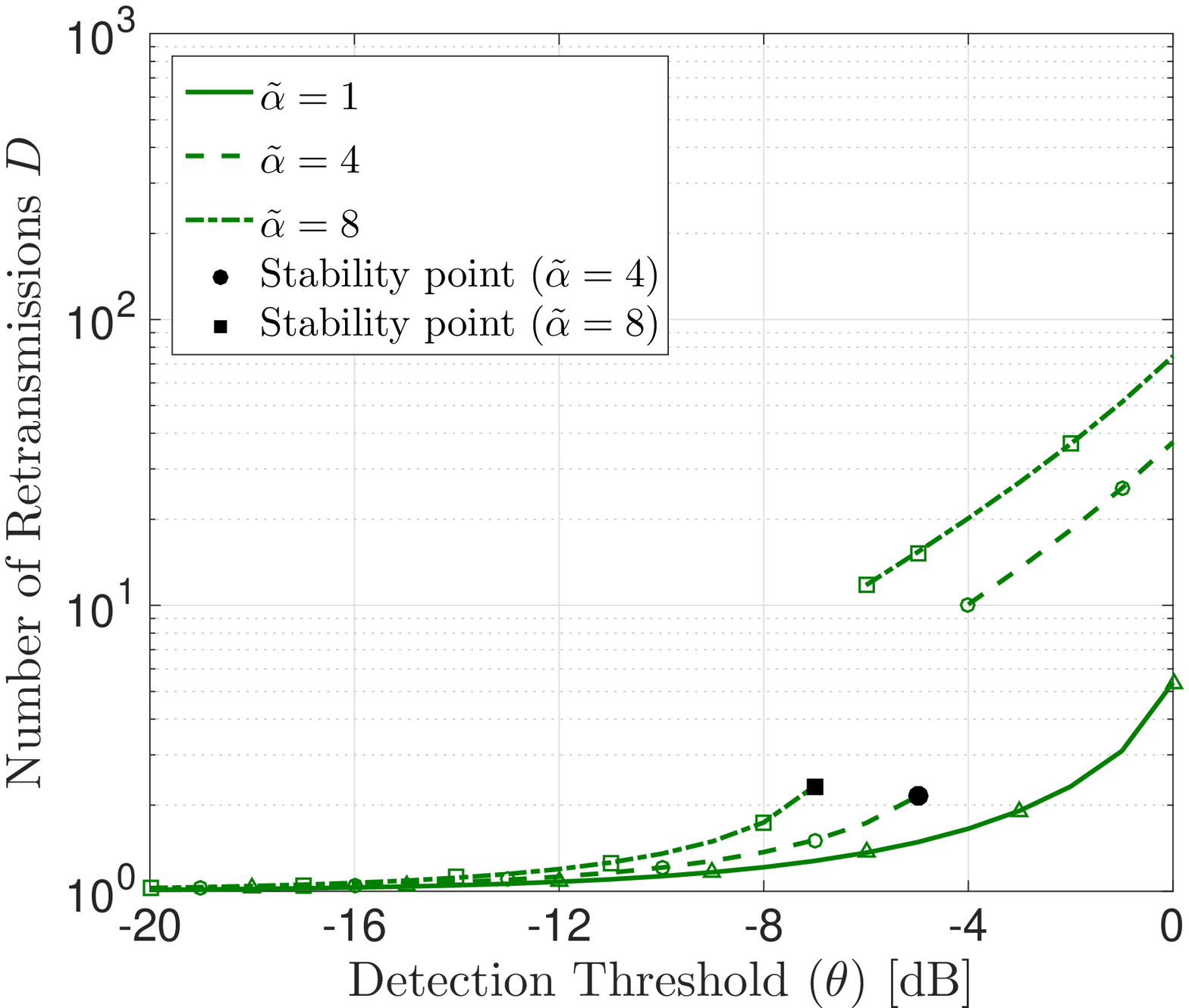}}
	\end{center}
	\vspace{-5mm}
	\caption{ Average number of retransmissions for the three schemes.}
	\label{fig:retransmissions}
	\vspace{-5mm}
 \hrulefill	
\end{figure*}

Fig.~\ref{fig:outage} shows the {spatially averaged  transmission success probabilities} at steady state versus the SINR detection threshold. First, we note the close match between the analysis and simulation results, which validates the developed mathematical framework. The discontinuities shown in Fig.~\ref{fig:outage} separate stable from unstable network operation. At stable network operation, the IoT devices are able to empty their buffers and go into idle state, which diminishes the aggregate interference. Increasing the required SINR threshold reduces the service rate drastically for two reasons. First, setting higher detection threshold reduces the probability of transmission success. Second, it increases the intensity of interfering deceives as the probability of flushing the queue is reduced, which reduces the average SINR. Consequently, $\theta$ has a composite effect on the service rate, and hence, the network stability is sensitive to $\theta$. Such sensitivity justifies the discontinuity point in the success probability curves in which a sudden transition occur from stable network operation (i.e., idle state probability $x_0>0$) to unstable network operation (i.e., $x_0=0$). Note that the magnitude of the discontinuity in each curve depends on the value of $x_0$ at which the transition occurs. For instance, the transition from stable to unstable network operation occurs in the power-ramping scheme for $\tilde{\alpha}=4$ and $\tilde{\alpha}=8$ at $x_0 = 0.71$ and $x_0=0.45$, respectively, which explains the higher discontinuity  gap for $\tilde{\alpha}=4$.


 Fig.~\ref{fig:queue_length} and Fig.~\ref{fig:waiting_time} show, respectively,  the average queue length $\mathbb{E}[Q_L]$ and  the average queueing delay $\mathbb{E}[W_q]$ for each of the transmission schemes at stable network operation. These two figures show an interesting behavior where the power-ramping scheme outperforms both the baseline and backoff schemes at low and moderate IoT  device density, $\tilde{\alpha}=1$ and $\tilde{\alpha}=4$. This is because the power-ramping scheme at these device densities offers a quicker buffer flushing before falling into instability which leads to a lower average queue length and average queueing delay. Moreover,  Fig.~\ref{fig:queue_length}  depicts that the power-ramping is able to extend the stability of the network at traffic arrival, $a=0.1$, for low and moderate IoT device intensities. However, at high IoT device intensity and traffic arrival $a=0.1$ the power-ramping offers the same stability performance but at higher average queue length. The behavior of the power-ramping scheme can be explained by its ability to prioritize the transmission of packets that experience previous transmission failures by increasing their power. Such prioritization works well for low device intensity that already imposes low probability of transmission failure at initial power control thresholds, and hence, only a few devices would ramp their powers. Increasing the intensity of IoT devices increases the interference and impose higher packet failure probabilities for low transmission powers, which increases the density of devices ramping their powers. In this case, packets are not successfully transmitted unless a high power threshold is reached, which enforces unnecessary phase transitions through ineffective power-ramping phases and imposes higher interference and delay. Consequently, at high traffic density, it is better to employ the baseline or backoff schemes which gives an equal probability for all packets to be successfully transmitted. Hence, increasing the probability of faster queue flushing, which increases the probability of being idle and decreases interference.  

 Fig.~\ref{fig:retransmissions} shows the average service (i.e., transmission) delay $\mathcal{D}$ for each of the transmission schemes for stable and unstable network. Fig.~\ref{fig:retransmissions} shows consistent insights with the previous results in which the power-ramping scheme works better for low and moderate device intensities  for both stable and unstable network. On the other hand, when the network is stable the baseline/backoff scheme works better for higher traffic intensities, and when the network is unstable the bakoff outperforms the baseline and the power-ramping at high IoT device intensity. Furthermore, Fig.~\ref{fig:retransmissions} sheds light on a reason for network instability in which a sudden transition occurs in the transmission delay that leads to service probability less than the arrival probability. 

All the pervious figures show an equivalent performance between the baseline and backoff schemes as long as the network is stable. Such behavior can be explained by looking into the optimum values for the backoff parameters $N$ and $q$, which are reported in Table \ref{Modes_Table}. It can be inferred from Table \ref{Modes_Table} along with Figs.~\ref{fig:outage}-\ref{fig:retransmissions} that whenever the network is stable the optimal parameters are $N=0$ and $q=1$, which reduces the backoff scheme to the baseline scheme. This explains the matching behavior between both schemes in Figs.~\ref{fig:outage}-\ref{fig:retransmissions} and shows that the backoff has only a potential in congested network scenarios. Stable network operation does not need backoff regulation for the channel access as the service rate is already  higher than the arrival rate. Consequently, employing the baseline scheme leads to quicker buffer flushing, which automatically relieves interference due to idle devices. On the other hand, at congested  network operation when all the devices maintain non-empty buffers, i.e. the network is unstable, it is mandatory to regulate the channel access through the appropriate backoff strategy which makes the backoff scheme outperforms the baseline as well as the power-ramping scheme.

\vspace{-1mm}
  \begin{table}[h!]
\centering
\caption{{ Optimum Values for Backoff Parameters.  }}
\vspace{-3mm}
\resizebox{0.4 \textwidth}{!}{\begin{tabular}{|c|c|c|c|c|c|}
\hline
\rowcolor{lightgray}\multirow{-2}{*}{}
& & & & & \\

\rowcolor{lightgray}\multirow{-2}{*}{ Intensity (\textbf{$\tilde{\alpha}$})}&  \multirow{-2}{*}{SINR thresholds (\textbf{$\theta$})}&  \multirow{-2}{*}{\# backoff slots (\textbf{$N$})}&  \multirow{-2}{*}{ backoff probability  (\textbf{$q$})} & \multirow{-2}{*}{ Mean Backoff time} &  \multirow{-2}{*}{ Stability}  \\ \hline 
\hline

 \multirow{2}{*}{\textbf{$1 $\; }} & \multirow{2}{*}{$[-10, -6, -2]$ dB} &  \multirow{2}{*}{$ 0  $} & \multirow{2}{*}{$1$}& \multirow{2}{*}{$0$}  & \multirow{2}{*}{Stable}\\
 & &  & & & \\ \cline{2-6}  \hline
 \hline

  \multirow{4}{*}{\textbf{$4 $\; }} & \multirow{2}{*}{$[-10, -6]$ dB} &  \multirow{2}{*}{$ 0  $} & \multirow{2}{*}{$1$} & \multirow{2}{*}{$0$} & \multirow{2}{*}{Stable} \\
 & & & & &\\ \cline{2-6} & \multirow{2}{*}{$-2$ dB} &  \multirow{2}{*}{$ 2 $}  & \multirow{2}{*}{$ .91 $} & \multirow{2}{*}{$3.1$}  & \multirow{2}{*}{Unstable}\\
 & & & & & \\   \hline
  \hline

  \multirow{6}{*}{\textbf{$8$\; }} & \multirow{2}{*}{$-10$ dB} &  \multirow{2}{*}{$ 0  $} & \multirow{2}{*}{$1$}& \multirow{2}{*}{$0$} & \multirow{2}{*}{Stable} \\
& & &  & &\\ \cline{2-6}
 & \multirow{2}{*}{$-6$ dB} & \multirow{2}{*}{$ 2 $} &   \multirow{2}{*}{$ .87 $}&  \multirow{2}{*}{$3.15$}&\multirow{2}{*}{Unstable}  \\
& & & & &\\  \cline{2-6}
 & \multirow{2}{*}{$-2$ dB} &  \multirow{2}{*}{$ 4 $}  & \multirow{2}{*}{$.29 $}& \multirow{2}{*}{$7.45$} & \multirow{2}{*}{Unstable}  \\
 & & & & & \\   \hline
\end{tabular}}
\label{Modes_Table}
\end{table}

\begin{figure}[t!]\label{fig:stability}
        \begin{center}
	  \subfigure[$\theta$ Dependent]{\includegraphics[width=3in]{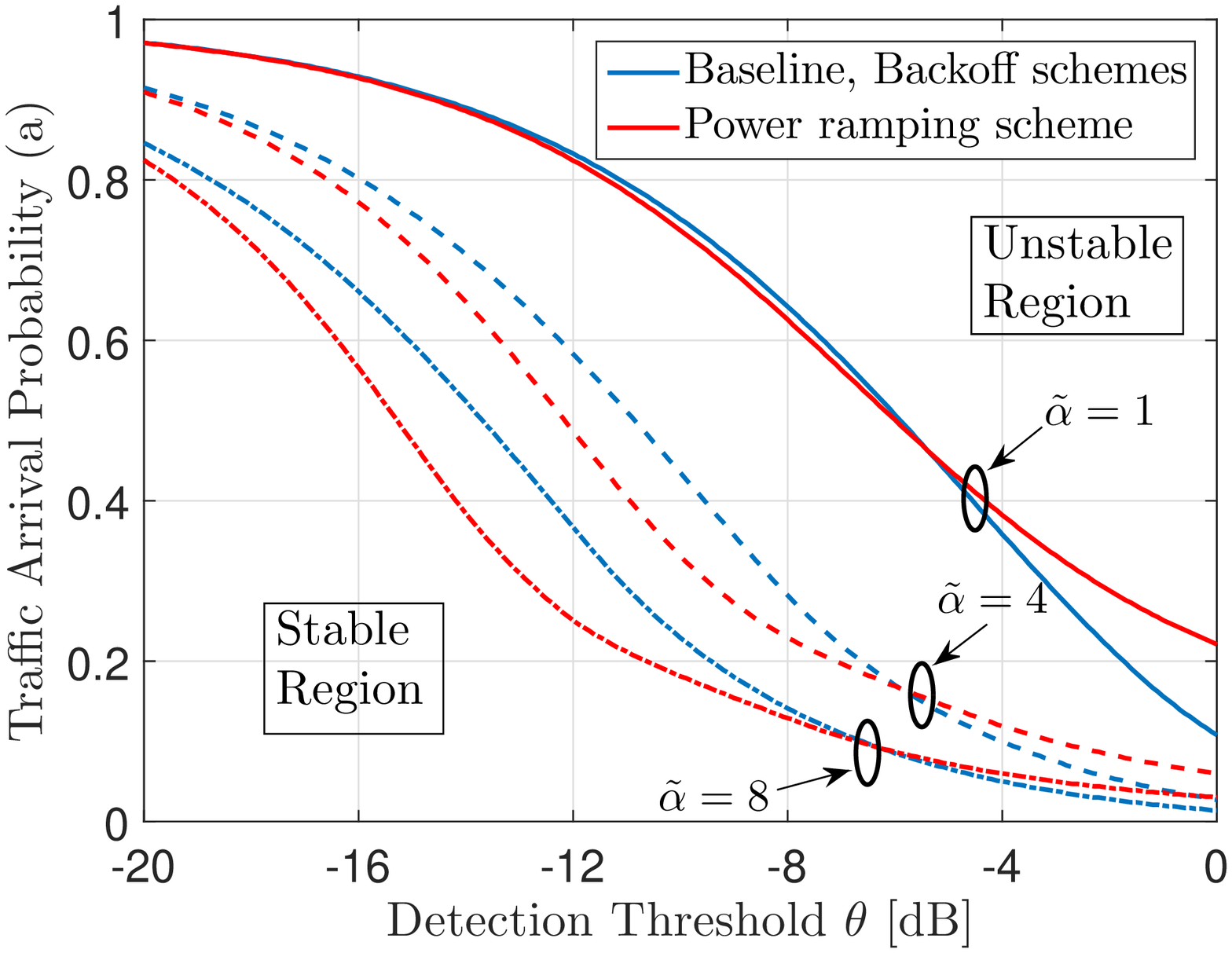}}
	  \subfigure[$\tilde{\alpha}$ Dependent]{\includegraphics[width=3in]{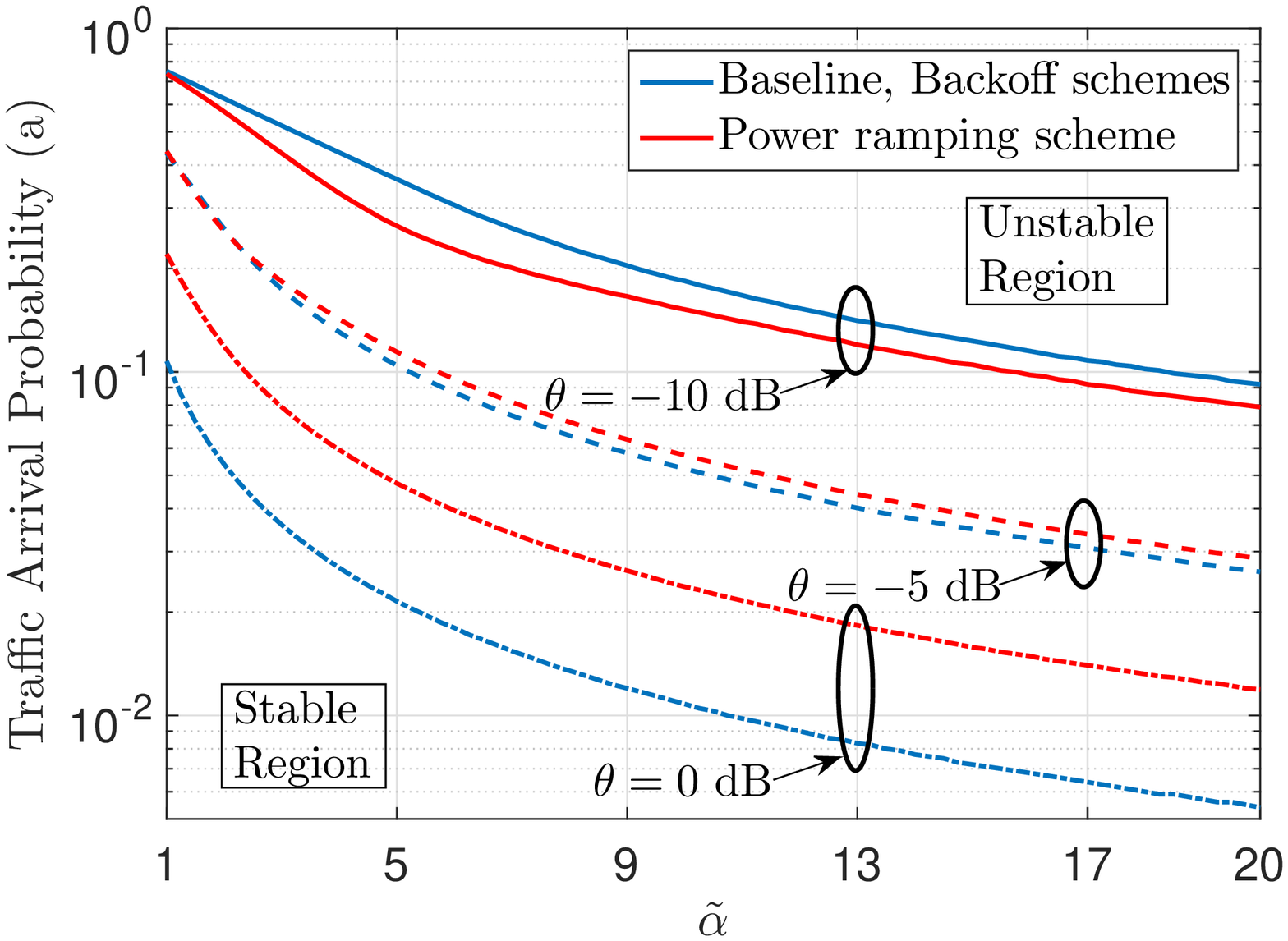}}
	\end{center}
	\vspace{-5mm}
	 \caption{Comparing the baseline, backoff, and power-ramping schemes stability regions boundaries.}
\label{fig:stability}
\end{figure}
%

Fig.~\ref{fig:stability} shows the scalability/stability tradeoff be plotting the Pareto-frontier of the stability regions in terms of the SINR detection threshold $\theta$, the packet arrival probability $a$, and device spatial intensity $\tilde{\alpha}$. The figure shows that the stability region shrinks for higher $\tilde{\alpha}$ and higher $\theta$, in which the cellular networks can only accommodate less traffic per IoT device.  As discussed earlier and shown in Table \ref{Modes_Table}, the baseline and backoff schemes are equivalent for stable network operation, and hence, the backoff scheme does not provide scalability gains for the IoT network. Fig.~\ref{fig:stability} shows that the power-ramping scheme provides a tangible extension for the stability region at high SINR, which is consistent with our findings in  Figs.~\ref{fig:outage}-\ref{fig:waiting_time}. On the other hand, the baseline scheme provides larger stability region for low SINR. At moderate SINRs both techniques have approximately similar performance with a slight advantage of the power-ramping scheme. Such advantage is because the power-ramping scheme assigns lower power to packets that experience fewer failures, which increases the service probability for packets transmitted at higher power. On the other hand, the baseline  scheme has similar success probability for all packets, which impose poor performance at low service probability. 

The key take away message from the results is that stable network operation cannot be maintained when scaling up the IoT device intensity. Operating at a lower detection threshold  ($\theta$) and/or decreasing the traffic intensity are two solutions to retain stable network operation when increasing the IoT device intensity. Furthermore,  network densification and/or increasing the number of channels can compensate for increasing the IoT device intensity to maintain stable network operation. When the network is stable, power-ramping scheme performs well for high SINR threshold as well as for moderate IoT intensities. On the other hand, for low SINR threshold and high IoT intensities, persistent transmission is more effective to keep lower delay and higher network stability than power-ramping scheme to avoid unnecessary transitions throughout lower power phases. Last but not least, backoff is not necessary at stable network operation. For unstable network operation, the power-ramping still achieves lower average number of retransmissions at moderate IoT device intensity, but at high IoT device intensity the backoff scheme is necessary to enforce backoff time before reattempting to access in order to relief  congestion. Consequently, interference levels are relaxed, transmission success probability is increased while providing a better average number of retransmissions.

\vspace{-4mm}
\section{Conclusions}\label{Conclusions}
\vspace{-2mm}
Exploiting stochastic geometry and queueing theory, this paper develops a novel traffic-aware spatiotemporal mathematical model for IoT cellular networks. The developed model is used to characterize the uplink scalability to serve massive IoT devices as well as to study and compare three transmission strategies employed by the IoT devices, which incorporate a combination of transmission persistency, backoff, and power-ramping. The proposed analysis abstracts the IoT devices and their serving base stations to a network of interacting queues where the interactions reside in the mutual interference between the devices. To this end, a two-dimensional discrete time Markov chain (DTMC) is proposed to model the queue and protocol states of each IoT devices where the transition matrices are populated with the transmission success probabilities obtained via stochastic geometry analysis. Also the stochastic geometry analysis depends on the steady state solution obtained by solving the DTMC.  Such interdependence between the DTMC solution and stochastic geometry analysis creates a causality problem that is solved via an iterative solution. The results characterize the scalability of uplink cellular networks by plotting the Pareto-frontier of the stability region, which defines the network parameters at which the cellular networks can serve all packets generated by the IoT devices in finite time. Moreover, the results showed different solutions to maintain stable network operation when scaling the IoT device intensity. Design insights for the employed transmission schemes are provided and the effective operation scenario of each scheme is highlighted. For instance, it is shown that the power-ramping scheme is effective at high SINR threshold and persistent fixed power transmission is effective at low SINR threshold. Last but not least, deferring transmission upon failure is not required at stable network operation.

\vspace{-2mm}
\appendix
\vspace{-1mm}
\subsection{Proof of Proposition 1 }\label{Correlation}
Let  $\tilde{\Phi}_{\rm t_i} \subset \Phi$ and $\mathcal{I}_{\rm t_i} \in \mathbb{R}$ denote the realization of the locations of the interfering IoT devices and the generated aggregate interference power at the test BS, respectively, at the $i^{th}$ time slot. The set $\tilde{\Phi}_{\rm t_i}$ is obtained from $ \Phi$ by independent thinning with probability $\mathcal{K}$, and hence, $\tilde{\Phi}_{\rm t_i}$ is a PPP $\forall i$. Also, $\tilde{\Phi}_{\rm t_i}$ and $\tilde{\Phi}_{\rm t_j}$ may have common elements for $i \neq j$. This appendix derives the joint transmission failure probability defined as
\vspace{-.1mm}
\small
\begin{align}\label{eq:correlation1} 
 \mathbb{P}\left\{\rm{SINR}_{\rm t_1} < \theta , \rm{SINR}_{\rm t_2} < \theta \right\}= 1-  \mathbb{P}  \underset{\RN{1}}{\underbrace{\left\{\rm{SINR}_{\rm t_1} >\theta \right\}}} - \mathbb{P} \underset{\RN{2}}{\underbrace{ \left\{\rm{SINR}_{\rm t_2} >\theta \right\}}}+ \underset{\RN{3}}{\underbrace{\mathbb{P}  \left\{\rm{SINR}_{\rm t_1} >\theta, \rm{SINR}_{\rm t_2} >\theta \right\}}},
\end{align}
\normalsize
where ${\rm SINR}_{\rm t_i}$ represents the SINR for the $i^{th}$ time slot. The terms $\RN{1}$ and $\RN{2}$ are obtained using \eqref{eq:Out1} by replacing $\bar{x}_0$ with $\mathcal{K}$. The term $\RN{3}$ can be defined as 
\vspace{.1mm}
\small
\begin{align} 
 \mathbb{P} \left\{ \frac{\rho h_{{\rm t_1}}}{\sigma^2 + \mathcal{I}_{\rm t_1}}\!\!>\!\!\theta,\frac{\rho h_{{\rm t_2}}}{\sigma^2 + \mathcal{I}_{\rm t_2}}\!\!>\!\!\theta \right\} {=} \exp\left\{- \frac{2\sigma^2 \theta}{\rho} \right\} \mathcal{L}_{\mathcal{I}_{\rm t_1}^{(\text{in})}, \mathcal{I}_{\rm t_2}^{(\text{in})}} \left(\frac{\theta}{\rho},\frac{\theta}{\rho} \right)\mathcal{L}_{\mathcal{I}_{\rm t_1}^{(\text{out})}, \mathcal{I}_{\rm t_2}^{(\text{out})}}\left(\frac{\theta}{\rho},\frac{\theta}{\rho} \right),\label{SINR_corr}
\end{align}
\normalsize

\noindent which is obtained by decomposing $\mathcal{I}_{\rm t_i}$ to $\mathcal{I}^{(\text{in})}_{\rm t_i} + \mathcal{I}^{(\text{out})}_{\rm t_i}$, where $\mathcal{I}^{(\text{in})}_{\rm t_i}$ and $\mathcal{I}^{(\text{out})}_{\rm t_i}$ denote the intra-cell and inter-cell interference, respectively, at time slot $t_i$. Then \eqref{SINR_corr} follows from the i.i.d. exponential distribution of $h_{\rm t_i}$ $\forall i$  along with the independence between $\mathcal{I}^{(\text{in})}_{\rm t_i} $ and  $\mathcal{I}^{(\text{out})}_{\rm t_i}$. Hence, correlations exist between  $\mathcal{I}^{(\text{in})}_{\rm t_i} $ and  $\mathcal{I}^{(\text{in})}_{\rm t_j} $ as well as  $\mathcal{I}^{(\text{out})}_{\rm t_i} $ and $\mathcal{I}^{(\text{out})}_{\rm t_j} $ due to the probability of common interferers. The joint Laplace transform between the inter-cell interferences in two time slots is expressed in \eqref{coor_h}, which is obtained using similar analogy to \cite{7067349} for the joint LT and \cite{elsawy2014stochastic} for the uplink scenario, in which $\mathcal{K} (1-\mathcal{K}) \mathcal{U}$ is the intensity of  $\{\Phi_{\rm t_i} \setminus \Phi_{\rm t_j} \}$ and   $\mathcal{K}^2 \mathcal{U}$ is the intensity of $\{\Phi_{\rm t_i} \cap \Phi_{\rm t_j} \} $. 
Let $V$ denote that Voronoi cell of the test BS and $|\cdot|$ denote the set cardinality. Also, let $K = \left| V \cap \{\Phi_1 \setminus \Phi_2 \} \right|, L = \left| V \cap \{\Phi_2 \setminus \Phi_1 \} \right|$, and $N = \left| V \cap \{\Phi_1 \cap \Phi_2 \}\right| $, then the conditional joint LT of the intra-cell interference is given by

\vspace{-2mm}
\small
\begin{align}\label{eq:corr}
&\mathcal{L}_{\mathcal{I}_{\rm t_1}^{(\text{in})}, \mathcal{I}_{\rm t_2}^{(\text{in})}}(s_1,s_2\mid K,L,N)= \mathbb{E}\left[e^{-s_1\!\! \underset{i \in V \cap \Phi_{\rm t_1}}{\sum} \!\!\!\!\rho g_{\rm t_1,i} -  s_2 \!\! \underset{j \in V \cap \Phi_{\rm t_2}}{\sum}\!\!\!\! \rho g_{\rm t_2,j} }\right] 
 = \frac{1}{(1+s_1\rho)^{K+N}(1+s_2\rho)^{L+N}}.
\end{align}
\normalsize

\begin{figure*}
\scriptsize
\begin{align}
&\mathcal{L}_{\mathcal{I}_{\rm t_1}^{(\text{out})}, \mathcal{I}_{\rm t_2}^{(\text{out})}}= \exp\left\{-2\pi \mathcal{K} \mathcal{U} \mathbb{E}_{P}\left[P^{\frac{2}{\eta}}    \right]   \left(   \mathcal{K} \!\!\!\! \int\limits_{(\rho)^{\frac{-1}{\eta}}}^{\infty}\!\!\!\!  \left(1- \frac{1}{1+s_1 y^\eta }\frac{1}{1+s_2 y^\eta }\right) dy    +  (1-\mathcal{K}) \left[  s_1^{\frac{2}{\eta}}\!\!\!\! \!\!\!\!  \int\limits_{(s_1\rho)^{\frac{-1}{\eta}}}^{\infty} \frac{y}{y^\eta +1} dy  +   s_2^{\frac{2}{\eta}}  \!\!\!\! \!\!\!\! \int\limits_{(s_2\rho)^{\frac{-1}{\eta}}}^{\infty} \frac{y}{y^\eta +1} dy \right]\right) \right\}  \label{coor_h}
\end{align}
\normalsize
\hrulefill
\end{figure*}

Putting all together, the joint failure probability in \eqref{eq:correlation1} can be expressed as in \eqref{eq:corr1}, where $\psi (N,\eta, \theta)= {}_2F_1\left(1,N-2/\eta,N+1-2/\eta,-\theta\right)$, $\varphi(\tau,\xi)= \left(  1+\frac{\tau \xi \tilde{\alpha}}{c}\right)^{-c}$ and $c=3.575$. Note that \eqref{eq:corr1} follows by substituting \eqref{eq:Out1}, \eqref{coor_h}, and \eqref{eq:corr} in \eqref{eq:correlation1}, averaging over $K$, $L$, and $N$, with are i.i.d. distributed with CDF in \cite[Eq.~(8)]{6576413}, and substituting  $\mathbb{E}_{P}\left[P^{\frac{2}{\eta}}    \right]  = \frac{\rho^{\frac{2}{\eta}}}{\pi \lambda}$ \cite{elsawy2014stochastic}. Finally, the Fig.~\ref{fig:correlation} in Proposition 1 is plotted using 

\vspace{-3mm}  
\small
\begin{align}
\mathbb{P} \left\{SINR_{\rm t_2} \!\!\!< \! \theta\! \mid \! SINR_{\rm t_1}\!\!\!  <\! \theta \right\}=\frac{\mathbb{P} \left\{SINR_{\rm t_1} \!\!\! <\! \theta, SINR_{\rm t_2} \!\!\! < \! \theta \right\} }{ \mathbb{P}  \left\{SINR_{\rm t_1} <\theta \right\} }.
\end{align}
\normalsize
\vspace{-3mm}  

 
 
\vspace{-1.5mm}
\begin{figure*}
\scriptsize
\begin{center}
\begin{align}\label{eq:corr1}
\mathbb{P}\left\{\rm{SINR}_{\rm t_1} < \theta , \rm{SINR}_{\rm t_2} < \theta \right\} &= 1-2 \varphi(1) \exp\left\{- \frac{\sigma^2 \theta}{\rho} -  \frac{2 \theta \tilde{\mathcal{U}}}{(\eta-2) \lambda } \;\psi (1,\eta, \theta) \right\} + \varphi(\mathcal{K} ) \; \varphi(1- \mathcal{K} )^2 \notag\\ 
& \! \!\!\!\!\!\!\!\!\!\!\!\!\!\!\!\!\!\!\!\!\!\!\!\!\!\!\!\!\!\!\!\!\!\times \exp\left\{- \frac{2 \sigma^2 \theta}{\rho} -  \frac{4 \theta \tilde{\mathcal{U}}}{\lambda } \left( \frac{1-\mathcal{K}}{\eta-2 }  \;\psi (1,\eta, \theta)  +  \frac{2 \mathcal{K} }{  \eta } \;\left[\frac{2+\theta}{4(1+\theta)} + \frac{(\eta-2) \theta \; \psi (2,\eta, \theta)}{8(\eta-1)}+\frac{\psi (1,\eta, \theta)}{\eta-2}\right] \right) \right\}
\end{align}
\end{center}
\hrulefill
\end{figure*}
\normalsize


\vspace{-5mm}
\subsection{Proof of Lemma \ref{lem:out_base}}\label{sec:AppA}
Due to the unscheduled transmission there are two source of interference, namely, an intra-cell and  an inter-cell interferences. To evaluate the interference experienced by a device, we find the LT of the aggregate intra-cell interference along with the inter-cell interference. Because of the independtancy of the PPP in diffrent regions \cite{ martin_book}, \eqref{sinr} can be written as follows:
\vspace{-.1mm}
\small
\begin{align}
p & = \exp\left\{- \frac{\sigma^2 \theta}{\rho} \right\} \mathcal{L}_{\mathcal{I}_{\text{out}}} \left(\frac{\theta}{\rho} \right) \mathcal{L}_{\mathcal{I}_{\text{in}}} \left(\frac{\theta}{\rho} \right).
\label{SINR1}
\end{align}
\normalsize
 Note that the nearest BS association and the employed power control enforce the following two conditions; (i) the intra-cell interference from an interfering device is equal to $\rho$, and (ii) the inter-cell interference from any interfering device is strictly less that $\rho$. Approximating the set of interfering devices by a PPP with independent transmit powers, the aggregate inter-cell interference received at the BS is obtained as:
\vspace{-1mm}
\small
\begin{align}\label{eq:AppA_3}
\mathcal{I}_{\text{out}}=  \sum\limits_{u_i\in \Phi \setminus \{o\} } \mathbbm{1}_{\{P_{i} \parallel u_i\parallel^{-\eta}<\rho\}}P_{i} g_i \parallel u_i\parallel^{-\eta} .
\end{align}
\normalsize
Ignoring the correlations between the transmission powers of the devices in the same and adjacent Voronoi cells, the Laplace Transform of \eqref{eq:AppA_3} can be approximated as:
\small
\begin{align}
&\mathcal{L}_{\mathcal{I}_{\text{out}}}(s) \approx \exp\left(-2\pi \; \bar{x_\circ}\; \tilde{\mathcal{U}} \; s^{\frac{2}{\eta}} \;\mathbb{E}_{P}\left[P^{\frac{2}{\eta}} \;   \right] \int_{(s\rho)^{\frac{-1}{\eta}}}^{\infty} \frac{y}{y^\eta +1} dy  \right).
\end{align}
\normalsize

\noindent  The LT is obtained by using the probability generating function (PGFL) of the PPP \cite{ martin_book} and following \cite{elsawy2014stochastic,7067349}, where  $\mathbb{E}_{x} [.]$ is the expectation with respect to the random variable $x$ , the Laplace Transform is obtained by substituting the value of $\mathbb{E}_{P}\left[P^{\frac{2}{\eta}}\right]$ from [Lemma 1,\cite{elsawy2014stochastic}].
The Intra-cell interference conditioned on the number of neighbors is given by:
\vspace{-2mm}
\begin{align}\label{eq:AppA_1}
\small
\mathcal{I}_{\text{in}\mid n} = \displaystyle{\sum\limits_{n} \rho g_n}.
\normalsize
\end{align}
The Laplace Transform of \eqref{eq:AppA_1} is obtained as:
\vspace{-3mm}
\begin{align}
\small
\mathcal{L}_{\mathcal{I}_{\text{in}}\mid n}(s)&= \mathbb{E}[e^{-s\mathcal{I}}] = \frac{1}{(1+s\rho)^n}.
										\normalsize
\end{align}
The probability mass function of the number of neighbors  $\mathcal{N}$ which is found in \cite{6576413} as: 
	\begin{align}\label{pdf_users}
	\mathbb{P}\{\mathcal{N} = n\} \approx \frac{\Gamma(n+c)}{\Gamma(n+1)\Gamma(c)} \frac{\mathcal{U}^n (\lambda c)^c}{(\mathcal{U}+\lambda c)^{n+c}}
	\end{align}
where $\Gamma(.)$ indicates the Gamma function, $c=3.575$ is a constant related to the approximate the PDF of the PPP Voronoi cell area in $\mathbb{R}^2$. Considering that there is only Inter-cell interference when the number of neighbors in the cell is 0, and both of inter-cell and intra-cell interference otherwise we can write equation~(\ref{SINR1}) as follows:
\vspace{-.1mm}
\small
\begin{align}\label{eq:AppA_2}
p= \exp\left\{- \frac{\sigma^2 \theta}{\rho} \right\} \mathcal{L}_{\mathcal{I}_{\text{out}}} \left(\frac{\theta}{\rho} \right)\left[ \mathbb{P} \left\{\mathcal{N}=0 \right\}+{\sum\limits_{n=1}^{\infty}\frac {\mathbb{P} \left\{\mathcal{N}=n \right\}}{(1+s\rho)^n}} \right].
\end{align}
\normalsize
After some manipulations, \eqref{eq:Out1} in Lemma 1 is obtained.

\vspace{-3mm}
\subsection{Proof of Lemma \ref{lem:out_ramp}}\label{sec:AppB}
\vspace{-2mm}
The intra-cell interference in this case is $\mathcal{I}_{\text{in}}={ \sum_{k=1}^{M}\sum_{n=1}^{N} \rho_k g_{n,k}}$, while the inter-cell interference is $\mathcal{I}_{\text{out}}={ \sum_{k=1}^{M} \; \sum_{u_i\in \Phi \setminus \{o\}} \mathbbm{1}_{\{P_{ik} \parallel u_i\parallel^{-\eta}<\rho_k\}}P_{ik} g_i \parallel u_i\parallel^{-\eta} }$. Hence, \eqref{sinr_ramp} can be written as 
\vspace{-3mm}
\begin{align}
\small
p_m =\exp\left\{- \frac{\sigma^2 \theta}{\rho_m} \right\} \prod\limits_{k=1}^{M} \mathcal{L}_{\mathcal{I}^{(k)}_{\text{out}}} \left(\frac{\theta}{\rho_m} \right) \mathcal{L}_{\mathcal{I}^{(k)}_{\text{in}}} \left(\frac{\theta}{\rho_m} \right).
\normalsize
 \label{SINR2}
\end{align}
Using similar procedure to the proof of Lemma \ref{lem:out_base}, \eqref{eq:Out_ramp} in Lemma  \ref{lem:out_ramp} is obtained. Note that \eqref{eq:Out_ramp} is an approximation because $\mathcal{L}_{\mathcal{I}^{(k)}_{\text{in}}}(\cdot)$ in \eqref{SINR2} is obtained by ignoring the correlations between the transmission powers of the devices in the same and adjacent Voronoi cells.


\bibliographystyle{IEEEtran}
\bibliography{IEEEabrv,StringDefinitions,refrences}
\vfill
\end{document}